\documentclass[journal,twoside,final]{IEEEtran}
\usepackage{amsfonts,amsmath,amssymb,bm}
\usepackage[ruled,linesnumbered,boxed]{algorithm2e}
\usepackage{psfrag,graphicx,epsfig,cite,enumitem}
\usepackage{color}
\usepackage{subfigure}
\usepackage[percent]{overpic}
\usepackage{multirow}
\usepackage{autobreak}
\usepackage{amsthm}
\usepackage{makecell}

\newtheorem{lemma}{Lemma}

\def \beqi{\begin{IEEEeqnarray}{rcl}\IEEEyesnumber}
\def \eeqi{\end{IEEEeqnarray}}

\def \bmat{\begin{bmatrix}}
\def \emat{\end{bmatrix}}

\DeclareMathOperator{\Tr}{Tr}
\begin{document}

\title{Active Reconfigurable Intelligent Surface Enhanced Spectrum Sensing for Cognitive Radio Networks}

\author{Jungang Ge, Ying-Chang Liang, \IEEEmembership{Fellow,~IEEE}, Sumei Sun, \IEEEmembership{Fellow,~IEEE}, \\Yonghong Zeng, \IEEEmembership{Fellow,~IEEE}, and Zhidong Bai

\thanks{This work has been submitted to the IEEE for possible publication. Copyright may be transferred without notice, after which  this version may no longer be accessible.}
\thanks{Part of this work was presented in IEEE GLOBECOM 2023 \cite{ge2023active}.}
\thanks{J. Ge is with the National Key Laboratory of Wireless Communications, and the Center for Intelligent Networking and Communications (CINC), University of Electronic Science and Technology of China (UESTC), Chengdu 611731, China (e-mail: {gejungang@std.uestc.edu.cn}).}
\thanks{Y.-C. Liang is with the Center for Intelligent Networking and Communications, University of Electronic Science and Technology of China, Chengdu 611731, China (e-mail: liangyc@ieee.org).}
\thanks{S. Sun and Y. Zeng are with the Institute for Infocomm Research, Agency for Science, Technology and Research, Singapore 138632 (e-mail: {sunsm@i2r.a-star.edu.sg; yhzeng@i2r.a-star.edu.sg}).}
\thanks{Z. Bai is with the Key Laboratory for Applied Statistics of MOE, School of Mathematics and Statistics, Northeast Normal University, Changchun 130024, China (e-mail: baizd@nenu.edu.cn).}
}

\maketitle
\begin{abstract}

In opportunistic cognitive radio networks, when the primary signal is very weak compared to the background noise, the secondary user requires long sensing time to achieve a reliable spectrum sensing performance, leading to little remaining time for the secondary transmission. To tackle this issue, we propose an active reconfigurable intelligent surface (RIS) assisted spectrum sensing system, where the received signal strength from the interested primary user can be enhanced and underlying interference within the background noise can be mitigated as well. In comparison with the passive RIS, the active RIS can not only adapt the phase shift of each reflecting element but also amplify the incident signals. Notably, we study the reflecting coefficient matrix (RCM) optimization problem to improve the detection probability given a maximum tolerable false alarm probability and limited sensing time. Then, we show that the formulated problem can be equivalently transformed to a weighted mean square error minimization problem using the principle of the well-known weighted minimum mean square error (WMMSE) algorithm, and an iterative optimization approach is proposed to obtain the optimal RCM. In addition, to fairly compare passive RIS and active RIS, we study the required power budget of the RIS to achieve a target detection probability under a special case where the direct links are neglected and the RIS-related channels are line-of-sight. Via extensive simulations, the effectiveness of the WMMSE-based RCM optimization approach is demonstrated. Furthermore, the results reveal that the active RIS can outperform the passive RIS when the underlying interference within the background noise is relatively weak, whereas the passive RIS performs better in strong interference scenarios because the same power budget can support a vast number of passive reflecting elements for interference mitigation.

\end{abstract}


\section{Introduction}\label{sec:intro}

The contradiction between the ever-increasing wireless traffics and limited radio spectrum resource has been identified as one of the main challenges in next-generation wireless networks \cite{you2021towards}. To address this problem, cognitive radio (CR), which allows the secondary user (SU) to access the spectrum of the primary user (PU), is proposed to improve the spectrum utilization efficiency \cite{liang2020dynamic}. In opportunistic cognitive radio networks, the SU should first detect the availability of the primary spectrum, and then it can start a secondary transmission if the primary spectrum is vacant. Therefore, spectrum sensing is a key enabling technology of opportunistic cognitive radio networks for its capability of detecting available vacant spectrum.

In general, spectrum sensing is fundamentally a hypothesis testing problem, which aims to identify two hypotheses, namely, $\mathcal{H}_0$: PU is absent and $\mathcal{H}_1$: PU is present. Various spectrum sensing methods have been studied, e.g., energy detection, cyclostationary detection, and coherent detection \cite{zeng2010review}. Besides, the eigenvalue-based detection methods are also developed to tackle practical issues such as noise uncertainty and correlated signals, and the sensing performance can be improved \cite{zeng2009eigenvalue}. In general, the performance of these methods essentially depends on the strength of received primary signals relative to the background noise and the sensing time. When the received primary signals at the SU are weak, the SU has to collect numerous signal samples to guarantee acceptable detection accuracy. However, the exploitation of the numerous signal samples means that long sensing time is required, which will inevitably lead to little remaining time for the secondary transmission. Therefore, to improve the spectrum sensing performance while minimizing the required sensing time as much as possible, the only way is enhancing the relative strength of the interested primary signals over the background noise. Remarkably, the background noise may contain the additive white Gaussian noise (AWGN) at the SU and the interference from other interferers transmitting in the same band \cite{ghasemi2008spectrum,zeng2009reliability,lin2020glrt}. The interferers may consist of malicious primary user emulation attackers \cite{chen2011cooperative} and other cognitive radio users, which do not need protection via spectrum sensing. With the fact that the impact of the AWGN can only be reduced by enhancing the strength of interested primary signals, mitigating the interference from other interferers becomes another possible way to further improve the spectrum sensing performance. However, the interference incurred by the interferers mainly depends on their transmit power and the channels to the SU, which are uncontrollable for the SU in conventional systems, making it almost infeasible to mitigate the interference.

Recently, the emerging reconfigurable intelligent surface (RIS) has been identified as a promising technology for 6G and beyond for its capability of reconstructing a programmable and intelligent wireless environment \cite{wu2019towards, basar2019wireless,liang2019large, liu2021reconfigurable,zhou2023assistance,zhang2024channel,chen2023transmission}. It actually acts as a tunable reflector of the wireless environment that can reflect different incident signal components in different manners. Besides, RIS can be introduced to various conventional wireless communication systems in an environment-friendly and compatible way, and the performance gains brought by deploying RISs have been widely investigated in physical layer secret communication system \cite{chen2019intelligent}, interference-nulling scenarios \cite{jiang2022interference}, non-orthogonal multiple access networks \cite{liu2022reconfigurable}, multi-cell networks \cite{pan2020multicell}, and so on.
In addition, RIS has also been introduced to spectrum sensing systems, making it possible to achieve a reliable sensing performance with less required sensing time \cite{ge2022ris, wu2021irs, lin2022intelligent, nasser2022intelligent, wu2023joint,ge2023ris}. In \cite{ge2022ris}, the authors investigate a RIS-assisted spectrum sensing system where a multi-antenna SU aims to detect the primary signals with maximum eigenvalue detection method, and the number of required reflecting elements (REs) to achieve a near-$1$ detection probability is analyzed by utilizing random matrix theory. The authors of \cite{wu2021irs} study a RIS-enhanced energy detection method, and the closed-form expression of the average detection probability is derived. In \cite{lin2022intelligent}, the authors design a RIS-aided spectrum sensing system where the RIS reflection dynamically changes according to a given codebook, and a weighted energy combination method is proposed to exploit the time-variant reflected signal power for spectrum sensing. In \cite{nasser2022intelligent}, the RIS's effect on sensing performance is investigated by comparing two typical configurations, namely, enhancing the SNR of the primary receiver and increasing the sensing SNR of SU, and the challenges for RIS-assisted spectrum sensing systems are also summarized. In \cite{wu2023joint}, the authors study an RIS-enhanced opportunistic CR system, where the RIS is configured to not only improve the sensing performance but also enhance the secondary transmission rate.

The above works mainly focus on the passive RIS-assisted spectrum sensing systems, where numerous REs are usually required to achieve a significant performance gain due to the double-fading attenuation of the cascaded channel, namely, the PU-RIS-SU channel. Although the power consumed by the control and phase shift switch circuits of each RE is small, it still requires considerable power consumption to support the basic operations of the REs. As a result, the maximum number of passive REs is limited by the power budget for the RIS, which also limits the maximum performance gain of the passive RIS. To tackle this issue, a new RIS architecture, which is referred to as active RIS, is proposed in \cite{long2021active}. In comparison with the passive RIS, the active RIS can not only adapt the phase shifts but also amplify the incident signals \cite{chen2023active}. Hence, the active RIS can realize a more significant performance gain with fewer REs. In particular, there are also some works investigating the active RIS-assisted spectrum sensing systems \cite{li2023active, xie2023enhancing}. In \cite{li2023active}, the authors investigate a spectrum sensing system assisted by multiple active RISs, where an energy efficiency maximization is studied by jointly optimizing the detection parameter and the reflecting coefficients matrices (RCMs) of the active RISs. In \cite{xie2023enhancing}, the authors compare the performance gains of the passive RIS and active RIS in the RIS-assisted spectrum sensing systems. Specifically, the detection probability maximization problems for both passive and active RIS are studied, and the number of REs required to achieve a near-$1$ detection probability is also analyzed. However, all the above works assume that the active RIS does not incur any background noise when the interested PU is not present, which is not practical due to the fact as follows. As the RIS is supposed to assist the sensing process no matter whether the PU is present or not, the active RIS will always forward the background noise even when the interested PU is absent. Besides, by fully leveraging the capability of the RIS, the interference component within the background noise can also be mitigated, leading to a further enhanced performance gain.

Motivated by the above considerations, we investigate a more general RIS-assisted spectrum sensing system consisting of an SU, a PU of interest to the SU, an active RIS, and multiple interferers. In particular, the SU is equipped with multiple antennas, and it aims to identify the presence of the primary signals via the maximum eigenvalue detection (MED) method \cite{zeng2008maximum}. The active RIS is deployed to assist the sensing process by enhancing the signals from the interested PU while suppressing that from the interferers. It is worth noting that the active RIS will always forward the background noise to SU, as it should work all the time to assist the sensing process. Consequently, the background noise components of the sensing signal samples are correlated. Our main contributions in this paper are summarized as follows.
\begin{itemize}
    \item We propose to first pre-whiten the received signal samples to deal with the underlying correlated background noise, such that the MED method can be performed to detect the PU's presence based on the whitened signal samples. Based on this setup, we consider optimizing the RCM to maximize the detection probability, given a maximum tolerable false alarm probability. Noting that the sample covariance matrix under $\mathcal{H}_1$ can be characterized as a spiked model from random matrix theory, this problem can be equivalently transformed to maximizing the largest eigenvalue of the population covariance matrix under $\mathcal{H}_1$.
    \item Given the intractability of the considered problem, we first transform it to a weighted mean square error minimization problem by leveraging the principle of the well-known WMMSE algorithm, which actually falls into the majorization-minimization framework. Then, an iterative optimization approach is developed to obtain the optimal solutions. Besides, the WMMSE-based approach can also be exploited to optimize the phase shift matrix for the passive RIS-assisted spectrum sensing system, while the step to obtain optimal phase shift matrix in each iteration should be modified with the semidefinite programming (SDP) technique to deal with the unit-modulus constraint on the passive REs.
    \item To fairly compare passive RIS and active RIS, we study the required power budget of the RIS to achieve a target detection probability under a special case where the direct links are neglected and the RIS-related channels are line-of-sight (LoS). Particularly, when the interferers are negligible, an optimal amplification factor can be observed for the active RIS, which mainly depends on the ratio of the power consumption to the input power of each active RE. When considering the impacts of the interferers, we show that the active RIS's reflecting coefficients can be configured with some heuristic receiver design principles for multi-antenna systems, e.g., matched-filter, zero-forcing, and minimum mean square error. Based on the closed-form expressions of RCM, the required power budget of the active RIS can be obtained and compared with that of the passive RIS. 
    \item Finally, we provide extensive simulations to demonstrate the effectiveness of the WMMSE-based reflecting coefficient optimization approaches for active and passive RIS-assisted spectrum sensing systems. In addition, the active RIS and passive RIS are comprehensively compared in terms of the required power budget to achieve a given target detection probability for different interference scenarios. Also, the performance gains brought by larger amplification factors and a larger number of REs are investigated. The results show that the active RIS can outperform the passive RIS when the underlying interference within the background noise is relatively weak, and the passive RIS performs better in strong interference scenarios.
\end{itemize}

The remainder of this paper is organized as follows. Section \ref{sec:aRIS4SS} introduces the active RIS-assisted spectrum sensing system, including the signal model, the adopted MED method, and the power consumption model of the active RIS. In Section \ref{sec:problem}, an optimization problem maximizing the largest eigenvalue of the population covariance matrix under $\mathcal{H}_1$ is formulated to improve the detection probability given a maximum tolerable false alarm probability. Then, Section \ref{sec:RCM_optimization} presents the WMMSE-based RCM optimization approach for the active RIS and a modified version for the passive RIS. Section \ref{sec:activeVSpassive} studies a special scenario with neglected direct links and LoS RIS-related channels, and the required power budget to achieve a target detection probability is derived for both active RIS and passive RIS for comparison. In Section \ref{sec:sim}, we provide extensive numerical simulations to evaluate the WMMSE-based RCM optimization approach, and the active RIS and passive RIS are comprehensively compared in terms of the required power budget to achieve a target detection probability for different interference scenarios. Finally, Section \ref{sec:conclusion} concludes this paper.

\emph{Notations:} The major notations used in this paper are as follows. The scalars, column vectors, and matrices are denoted by lowercase, bold lowercase, and bold uppercase symbols (e.g., $x$, $\mathbf{x}$, and $\mathbf{X}$), respectively. The Hermitian of $\mathbf{x}$ and $\mathbf{X}$ are respectively denoted by $\mathbf{x}^H$ and $\mathbf{X}^H$. $\Tr(\cdot)$ and ${\rm blkdiag}(\cdot)$ respectively denote the trace and block diagonalization operations. $|\cdot|$, $\|\cdot\|$, and $\|\cdot\|_{\infty}$ represent the absolute value, $L_2$-norm, and $L_{\infty}$-norm operations. $\mathbf{x}\sim\mathcal{CN}(\boldsymbol{\mu}, \boldsymbol{\Sigma})$ means that $\mathbf{x}$ is a complex Gaussian random vector with mean $\boldsymbol{\mu}$ and covariance matrix $\boldsymbol{\Sigma}$.

\section{Active RIS Assisted Spectrum Sensing System}\label{sec:aRIS4SS}

\begin{figure}[tbp]
    \begin{center}
        \epsfxsize=0.7\linewidth
        \epsffile{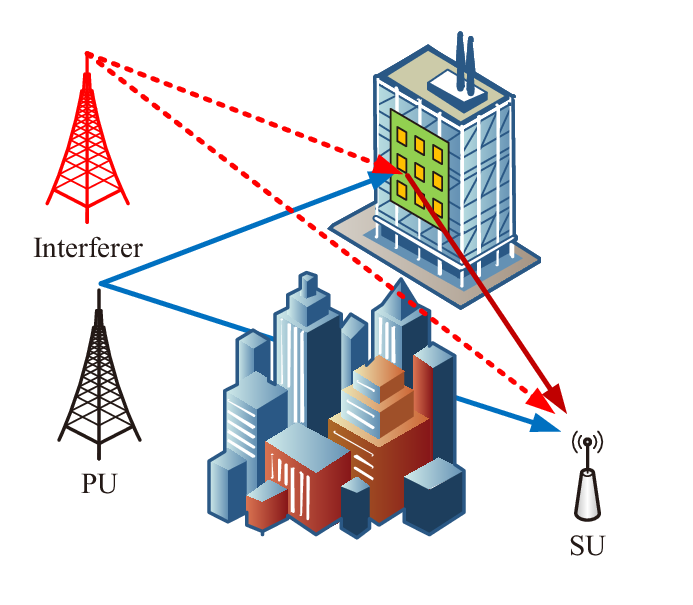}
    \end{center}
    \caption{The considered active RIS assisted spectrum sensing system.}\label{fig:sys_model}
\end{figure}

\subsection{Signal Model}\label{subsec:sig_model}

As shown in Fig. \ref{fig:sys_model}, we consider an active RIS-assisted spectrum sensing system, which consists of a single-antenna PU, $K$ single-antenna interferers, an active RIS with $M$ elements, and a SU equipped with $N$ antennas. Specifically, we use $\mathbf{d}_0\in\mathbb{C}^{N}$, $\mathbf{f}_0\in\mathbb{C}^{M}$ and $\mathbf{G}\in\mathbb{C}^{N\times M}$ to denote the PU-SU channel, PU-RIS channel and RIS-SU channel, respectively. Besides, the interferer-SU channel and interferer-RIS channel of the $k$-th interferer are represented by $\mathbf{d}_k$ and $\mathbf{f}_k$. In particular, we assume perfect channel state information knowledge of the above channels to investigate the upper-bound performance of the active RIS-assisted spectrum sensing system. Moreover, the RCM of the active RIS is a diagonal matrix denoted by $\mathbf{\Phi}={\rm diag}([\phi_1, \cdots, \phi_M])$ with $\phi_{m}=a_{m}e^{j\theta_m}$, where $a_{m}$ and $\theta_m$ are respectively the amplitude and the phase of $\phi_{m}$. Note that the active RIS will also forward the thermal noise because the active RIS should always work no matter whether the PU is active or not. Let $\mathbf{h}_0 \triangleq\mathbf{d}_0+\mathbf{G}\mathbf{\Phi}\mathbf{f}_0$ denote the equivalent channel between the PU and SU and $\mathbf{h}_k\triangleq\mathbf{d}_k+\mathbf{G}\mathbf{\Phi}\mathbf{f}_k$ denote the equivalent channel between interferer $k$ and SU, the $t$-th signal sample during a sensing interval under the two hypotheses can be respectively expressed as
\begin{align}
    &\hspace{-0.5em}\mathcal{H}_0\!:\! \mathbf{y}[t]=\sum_{k=1}^{K}\alpha_k\mathbf{h}_ks_k[t]+\mathbf{G\Phi}\mathbf{n}_{\rm R}[t]+\mathbf{n}_{\rm S}[t], \\
    &\hspace{-0.5em}\mathcal{H}_1\!:\! \mathbf{y}[t]=\mathbf{h}_0 s_0[t]+\sum_{k=1}^{K}\alpha_k\mathbf{h}_ks_k[t]+\mathbf{G\Phi}\mathbf{n}_{\rm R}[t]+ \mathbf{n}_{\rm S}[t],
\end{align}
where $s_0[t]\sim\mathcal{CN}(0, p_0)$ denotes the PU's transmitted symbol with $p_0$ the transmitting power, $s_k[t]\sim\mathcal{CN}(0, p_k)$ denotes the emitted signals of $k$-th interferer with $p_k$ its transmitting power, $\mathbf{n}_{\rm R}[t]\sim\mathcal{CN}(\mathbf{0}, \sigma_1^2\mathbf{I}_{M})$ is the thermal noise at the active RIS, $\mathbf{n}_{\rm S}[t]\sim\mathcal{CN}(\mathbf{0}, \sigma_2^2\mathbf{I}_{N})$ denotes the AWGN at SU, $\alpha_k$ is a binary indicator, and $\alpha_k=1$ if the $k$-th interferer is active and $\alpha_k=0$ otherwise. In addition, we use $\zeta_k={\rm Pr}(\alpha_k=1)$ to denote the probability of interferer $k$ being active.

\subsection{MED with Whitened Signal Samples}\label{subsec:MED}

Without loss of generality, we utilize the MED method to identify the presence of the primary signals. Let 
\begin{equation*}
    \tilde{\mathbf{z}}[t]=\sum_{k=1}^{K}\alpha_k\mathbf{h}_ks_k[t]+\mathbf{G\Phi}\mathbf{n}_{\rm R}[t]+\mathbf{n}_{\rm S}[t],
\end{equation*}
we have $\tilde{\mathbf{z}}[t]\sim\mathcal{CN}(0, \mathbf{R})$, where 
\begin{equation*}
    \mathbf{R}\triangleq\sum_{k=1}^{K}\zeta_kp_k\mathbf{h}_k\mathbf{h}_k^H+\sigma_2^2\mathbf{I}_N +\sigma_1^2\mathbf{G}\mathbf{\Phi}\mathbf{\Phi}^H\mathbf{G}
\end{equation*}
is the population covariance matrix of $\tilde{\mathbf{z}}[t]$. It can be observed that the interferers and the active RIS cause a correlated noise component in the sensing signal samples. To deal with this issue, we have to first pre-whiten the signal samples as follows. Let $\mathbf{Q}$ be the square root of $\mathbf{R}$ such that $\mathbf{R}=\mathbf{Q}^2$, the signal samples can be whitened by
\begin{equation*}
    \mathbf{x}[t] = \mathbf{Q}^{-1}\mathbf{y}[t].
\end{equation*}
Denoting $\mathbf{z}[t]=\mathbf{Q}^{-1}\tilde{\mathbf{z}}[t]\sim\mathcal{CN}(\mathbf{0}, \mathbf{I}_N)$, the whitened signal samples for the two hypotheses can be written as
\begin{align*}
    &\mathcal{H}_0: \mathbf{x}[t]=\mathbf{z}[t], \\
    &\mathcal{H}_1: \mathbf{x}[t]=\mathbf{Q}^{-1}\mathbf{h}_0 s_0[t]+ \mathbf{z}[t].
\end{align*}
Then, MED can be performed based on the whitened signal samples. The sensing capability of the MED method essentially originates from the difference between the population covariance matrices under the two hypotheses, i.e.,
\begin{align*}
    &\mathcal{H}_0: \mathbf{R}^0_{\mathbf{x}\mathbf{x}}=\mathbf{I}_{N}, \\
    &\mathcal{H}_1: \mathbf{R}^1_{\mathbf{x}\mathbf{x}}=p_0\mathbf{Q}^{-1}\mathbf{h}_0\mathbf{h}_0^H\mathbf{Q}^{-1}+ \mathbf{I}_{N}.
\end{align*}
Note that the largest eigenvalue of $\mathbf{R}^0_{\mathbf{x}\mathbf{x}}$ is $1$ and that of $\mathbf{R}^1_{\mathbf{x}\mathbf{x}}$ is $p_0\mathbf{h}_0^H\mathbf{R}^{-1}\mathbf{h}_0+1$, thus the presence of the primary signals can be detected by checking the largest eigenvalue of the covariance matrix. However, in practice, we can only obtain the sample covariance matrix
\begin{equation*}
    \hat{\mathbf{R}}_{\mathbf{x}\mathbf{x}} = \frac{1}{T} \sum_{t=1}^{T}\mathbf{x}[t]\mathbf{x}[t]^H,
\end{equation*}
which is usually not a good approximate of $\mathbf{R}_{\mathbf{x}\mathbf{x}}$ due to the limited number of signal samples. The relationship between the largest eigenvalue of $\hat{\mathbf{R}}_{\mathbf{x}\mathbf{x}}$ and that of $\mathbf{R}_{\mathbf{x}\mathbf{x}}$ is provided by random matrix theory \cite{ge2021large}. In particular, the largest eigenvalue of $\hat{\mathbf{R}}_{\mathbf{x}\mathbf{x}}$ under $\mathcal{H}_0$ can be characterized by the Tracy-Widom distribution \cite{ge2021large}, and the detection threshold $\gamma_{\rm th}$ for a given false alarm probability $\alpha$ can be calculated by
\begin{equation}\label{eq:detection_th}
    \gamma_{\rm th} = N^{-\frac{2}{3}}(1+\sqrt{c})^{\frac{4}{3}}\sqrt{c}F_2^{-1}(1-\alpha)+(1+\sqrt{c})^2,
\end{equation}
where $c=N/T$, and $F_2^{-1}(\cdot)$ is the quantile function of the Tracy-Widom distribution of order $2$. Finally, the presence of the primary signals can be detected by comparing the largest eigenvalue of $\hat{\mathbf{R}}_{\mathbf{x}\mathbf{x}}$ with $\gamma_{\rm th}$.

\subsection{Power Consumption Model of Active RIS}\label{subsec:consum_model}

In contrast to the passive RIS, the magnitudes of the reflecting coefficients of the active RIS can exceed $1$ for amplifying the incident signals. As a result, the active RIS also requires power for signal amplification, in addition to the power consumed by the control and phase shift switch circuits of the REs in passive RIS. Specifically, the power consumption of the active RIS can be modeled as \cite{long2021active}
\begin{equation}\label{eq:power_consume}
    P_{\rm ARIS} = M(P_{\rm C} + P_{\rm DC}) + P_{\rm out},
\end{equation}
where $P_{\rm C}$ is the power consumption for the control and phase shift switch circuits in each RE, $P_{\rm DC}$ denotes the direct current power consumption at each active RE, and $P_{\rm out}$ is the power of the amplified signals, which in the considered active RIS assisted spectrum sensing system is given by
\begin{equation*}
    P_{\rm out} = \sum_{k=0}^{K}\zeta_kp_k\|\mathbf{\Phi}\mathbf{f}_k\|^2+\sigma_1^2\|\mathbf{\Phi}\mathbf{1}_{M}\|^2,
\end{equation*}  
where we define $\zeta_0=1$ for ease of notation.

\section{Problem Formulation}\label{sec:problem}

To improve the performance of the active RIS-assisted spectrum sensing system, we can formulate a detection probability maximization problem given a maximum tolerable false alarm probability. From the viewpoint of random matrix theory, the largest eigenvalue of the sample covariance matrix can be characterized with the spiked model, and an exact separation phenomenon of the largest sample eigenvalue can be observed when the largest population eigenvalue, namely, $p_0\mathbf{h}_0^H\mathbf{R}^{-1}\mathbf{h}_0+1$, becomes larger \cite{couillet2011random}. By utilizing the exact separation phenomenon of the largest sample eigenvalue, we can derive that the detection probability can be optimized by maximizing $p_0\mathbf{h}_0^H\mathbf{R}^{-1}\mathbf{h}_0+1$. Recalling that $\mathbf{h}_0=\mathbf{d}+\mathbf{G}\mathbf{\Phi f}_0$, $\mathbf{R}=\sum_{k=1}^{K}\zeta_kp_k\mathbf{h}_k\mathbf{h}_k^H+\sigma_2^2\mathbf{I}_N +\sigma_1^2\mathbf{G}\mathbf{\Phi}\mathbf{\Phi}^H\mathbf{G}$, and $\mathbf{h}_k=\mathbf{d}_k+\mathbf{G}\mathbf{\Phi}\mathbf{f}_k$, we therefore consider an optimization problem as follows. 
\begin{subequations}
    \begin{align}
        \textbf{P1:} &\max_{\mathbf{\Phi}}\mathbf{h}_0^H\!\left(\sum_{k=1}^{K}\zeta_kp_k\mathbf{h}_k\mathbf{h}_k^H\!+\!\sigma_2^2\mathbf{I}_N \!+\!\sigma_1^2\mathbf{G}\mathbf{\Phi}\mathbf{\Phi}^H\mathbf{G}\right)^{\hspace{-0.2em}-1}\!\!\!\mathbf{h}_0 \nonumber \\
        \text{s.t.} &\sum_{k=0}^{K}\zeta_kp_k\|\mathbf{\Phi}\mathbf{f}_k\|^2\!+\!\sigma_1^2\|\mathbf{\Phi}\mathbf{1}_{M}\|^2\!\leq\! \bar{P}_{\rm out}, \label{eq:p1b}\\
        &a_{m} \leq a_{\max},\label{eq:p1c}
    \end{align} 
\end{subequations}
where $\bar{P}_{\rm out}\triangleq P_{\rm ARIS}-M(P_{\rm C} + P_{\rm DC})$ with $P_{\rm ARIS}$ the power budget of the active RIS, $\mathbf{1}_{M}$ is an $M$-dimensional column vector with all elements of $1$. Despite the convex constraints \eqref{eq:p1b} and \eqref{eq:p1c}, \textbf{P1} is quite difficult to tackle due to the non-convex objective.

\section{Reflecting Coefficient Matrix Optimization}\label{sec:RCM_optimization}

In this section, we first study the original optimization problem \textbf{P1} for active RIS, and we will show that the optimal RCM can be obtained by the majorization-minimization framework. Besides, this framework can also be exploited to optimize the phase shift matrix for the passive RIS-assisted spectrum sensing system with some minor modifications.

\subsection{WMMSE-based RCM Optimization for Active RIS}\label{subsec:WMMSE_active}

To introduce an equivalent weighted mean square error minimization problem of \textbf{P1}, we first show the principle of the well-known WMMSE framework \cite{zhao2023rethinking, shi2015secure}, which unfolds as the following lemma \cite{zhao2023rethinking}.
\begin{lemma}[Principle of the WMMSE framework]\label{lemma:wmmse}
    Given two matrices $\mathbf{A}\in\mathbb{C}^{n\times p}$, $\mathbf{B}\in\mathbb{C}^{p\times l}$, let $\mathbf{N}$ be a positive definite matrix, we have
    \begin{align}
        &\log\det(\mathbf{I}+\mathbf{A}\mathbf{B}\mathbf{B}^H\mathbf{A}^H\mathbf{N}^{-1}) \nonumber\\
        &\qquad = \max_{\mathbf{\Omega}>\mathbf{0}, \mathbf{\Gamma}}\log\det(\mathbf{\Omega})-\Tr(\mathbf{\Omega}\mathbf{E}(\mathbf{\Gamma}, \mathbf{B}))+l,\label{eq:wmmse_principle}
    \end{align}
    where $\mathbf{E}(\mathbf{\Gamma}, \mathbf{B})$ is defined as
    \begin{equation}
        \mathbf{E}(\mathbf{\Gamma}, \mathbf{B}) = (\mathbf{I-\mathbf{\Gamma}^H\mathbf{A}\mathbf{B}})(\mathbf{I-\mathbf{\Gamma}^H\mathbf{A}\mathbf{B}})^H+\mathbf{\Gamma}^H\mathbf{N}\mathbf{\Gamma},
    \end{equation}
    $\mathbf{\Gamma}\in\mathbb{C}^{n\times l}$ and $\mathbf{\Omega}\in\mathbb{C}^{l\times l}$ are two auxiliary matrices with $\mathbf{\Omega}$ a positive definite matrix. Moreover, the optimal $\mathbf{\Gamma}$ and $\mathbf{\Omega}$ for the right-hand side (RHS) of \eqref{eq:wmmse_principle} are respectively given by
    \begin{equation}\label{eq:opt_Gamma}
        \mathbf{\Gamma}_{\rm opt} = \left(\mathbf{N}+\mathbf{A}\mathbf{B}\mathbf{B}^H\mathbf{A}^H\right)^{-1}\mathbf{A}\mathbf{B},
    \end{equation}
    and
    \begin{equation}\label{eq:opt_Omega}
        \mathbf{\Omega}_{\rm opt} = \left(\mathbf{E}\left(\mathbf{\Gamma}_{\rm opt}, \mathbf{B}\right)\right)^{-1} = \left(\mathbf{I}-\mathbf{\Gamma}_{\rm opt}^{H}\mathbf{A}\mathbf{B}\right)^{-1}.
    \end{equation}
\end{lemma}

To obtain the optimal $\mathbf{\Phi}$ in \textbf{P1}, we have
\begin{align*}
    \arg\max_{\mathbf{\Phi}}\mathbf{h}_0^H\mathbf{R}^{-1}\mathbf{h}_0 
    &= \arg\max_{\mathbf{\Phi}}\log(1+\mathbf{h}_0^H\mathbf{R}^{-1}\mathbf{h}_0)\\
    &= \arg\max_{\mathbf{\Phi}}\log\det(\mathbf{I}+\mathbf{h}_0\mathbf{h}_0^H\mathbf{R}^{-1}),
\end{align*}
where the second equality comes from the fact that $\det(\mathbf{I}+\mathbf{AB})=\det(\mathbf{I}+\mathbf{BA})$. Therefore, using Lemma \ref{lemma:wmmse}, \textbf{P1} can be equivalently transformed into a weighted mean square error minimization problem as
\begin{subequations}
    \begin{align}
        \textbf{P2:} & \min_{\mathbf{\Phi}, \mathbf{u}, \omega} \omega\epsilon(\mathbf{u}, \mathbf{\Phi})-\log \omega \label{eq:p3a} \\
        & \text{s.t. } \eqref{eq:p1b}, \eqref{eq:p1c},
    \end{align} 
\end{subequations}
where $\mathbf{u}$, $\omega$ and $\epsilon(\mathbf{u}, \mathbf{\Phi})$ respectively play the same roles of $\mathbf{\Gamma}$, $\mathbf{E}(\mathbf{\Gamma}, \mathbf{B})$, and $\mathbf{\Omega}$ in Lemma \ref{lemma:wmmse}. To be specific, $\epsilon(\mathbf{u}, \mathbf{\Phi})$ is given by
\begin{align*}
    \epsilon(\mathbf{u}, \mathbf{\Phi})=p_0|\mathbf{u}^H\mathbf{h}_0-1|^2+
    \sum_{k=1}^{K}\zeta_kp_k|\mathbf{u}^H\mathbf{h}_k|^2\\
    +\sigma_1^2\|\mathbf{u}^H\mathbf{G}\mathbf{\Phi}\|^2+\sigma_2^2\|\mathbf{u}\|^2.
\end{align*}
In fact, \textbf{P2} falls into the majorization-minimization algorithmic framework, and $\mathbf{u}$, $\omega$ are exactly the parameters for constructing a surrogate function of the original objective function in \textbf{P1}. Note that \eqref{eq:p3a} is convex for each of the variables $\omega$, $\mathbf{u}$, $\mathbf{\Phi}$, \textbf{P2} can be solved by sequentially optimizing one of the three variables while keeping the other two fixed. With the first-order optimality condition, according to \eqref{eq:opt_Gamma} and \eqref{eq:opt_Omega}, the optimal $\omega$ and $\mathbf{u}$ are respectively given by
\begin{align}
    \!\!\!\mathbf{u}_{\rm opt} \!&=\!\! \left(\sum_{k=0}^{K}\zeta_kp_k\mathbf{h}_k\mathbf{h}_k^H\!\!+\!\sigma_1^2\mathbf{G}\mathbf{\Phi}\mathbf{\Phi}^H\mathbf{G}^H\!+\!\sigma_2^2\mathbf{I}_N\right)^{-1}\!\!\mathbf{h}_0.\label{eq:opt_u}\\
    \!\!\!\omega_{\rm opt} \!&=\! \frac{1}{\epsilon(\mathbf{u}, \mathbf{\Phi})}, \label{eq:opt_omega}
\end{align}
With fixed $\mathbf{u}$ and $\omega$, \textbf{P2} can be equivalently rewritten as the following optimization problem with respect to $\mathbf{\Phi}$.
\begin{subequations}
    \begin{align}
        \textbf{P2-1:} & \min_{\mathbf{\Phi}} \epsilon(\mathbf{u}_{\rm opt}, \mathbf{\Phi}) \label{eq:p3-1a} \\
        & \text{s.t. } \eqref{eq:p1b}, \eqref{eq:p1c}.
    \end{align}
\end{subequations}
Let $\mathbf{F}_k$ be a diagonal matrix consisting of all the elements of $\mathbf{f}_k$, i.e., $\mathbf{F}_k={\rm diag}(\mathbf{f}_k)$, $\forall k=0,1,\cdots, K$, and $\pmb{\phi}=[\phi_1, \cdots, \phi_M]^{T}$, we have
\begin{align*}
    &\mathbf{h}_k =\mathbf{d}_k+\mathbf{G}\mathbf{\Phi}\mathbf{f}_k = \mathbf{d}_k+\mathbf{G}\mathbf{F}_k\pmb{\phi} = \mathbf{A}_k\bar{\pmb{\phi}},\\
    &\|\mathbf{u}^H\mathbf{G}\mathbf{\Phi}\|^2 = |{\rm diag}(\mathbf{u}^H\mathbf{G})\pmb{\phi}|^2
    =|\mathbf{B}\bar{\pmb{\phi}}|^2,
\end{align*}
where $\mathbf{A}_k\triangleq[\mathbf{G}\mathbf{F}_k, \mathbf{d}_k]$, $\mathbf{B} \triangleq {\rm diag}([\mathbf{u}^H\mathbf{G}, 0])$ and $\bar{\pmb{\phi}}\triangleq[\pmb{\phi}^T, 1]^T$. Therefore, \textbf{P2-1} can be equivalently expressed as
\begin{subequations}
    \begin{align}
        \textbf{P2-2:} & \min_{\bar{\pmb{\phi}}} \bar{\epsilon}(\bar{\pmb{\phi}}) \label{eq:p3-2a} \\
        &\text{s.t. } \bar{\pmb{\phi}}^H\bar{\mathbf{J}}\bar{\pmb{\phi}}\leq \bar{P}_{\rm out},\\
        &\quad\  |\bar{\pmb{\phi}}_{m}|\leq a_{\max}, \forall m = 1, \cdots, M,\\
        &\quad\  \bar{\pmb{\phi}}_{M+1}=1,
    \end{align}
\end{subequations}
where 
\begin{equation*}
    \bar{\epsilon}(\bar{\pmb{\phi}})=p_0\left|\mathbf{u}^H\mathbf{A}_0\bar{\pmb{\phi}}-1\right|^2+\sum_{k=1}^{K}\zeta_kp_k\left|\mathbf{u}^H\mathbf{A}_k\bar{\pmb{\phi}}\right|^2+\sigma_1^2\|\mathbf{B}\bar{\pmb{\phi}}\|^2,
\end{equation*}
and
\begin{equation*}
    \mathbf{J}\!=\!{\rm diag}\!\left(\sum_{k=0}^{K}\zeta_kp_k|f_{k,1}|^2\!\!+\!\sigma_1^2, \cdots, \sum_{k=0}^{K}\zeta_kp_k|f_{k,M}|^2\!\!+\!\sigma_1^2, 0\right).
\end{equation*}
Noting that \textbf{P2-2} is a convex problem, we can obtain the optimal $\mathbf{\Phi}$ by solving \textbf{P2-2} with the convex optimization tools such as CVX \cite{grant2014cvx}. Finally, the WMMSE-based algorithm to solve \textbf{P1} is summarized in \textbf{Algorithm} \ref{Alg:WMMSE}.
\begin{algorithm}[hthp]
    \caption{The WMMSE-based algorithm for \textbf{P1}\label{Alg:WMMSE}}
    Initialize $\mathbf{\Phi}$ such that \eqref{eq:p1b} and \eqref{eq:p1c} are satisfied\;
    \Repeat{the convergence of $\omega$}{
        Update $\mathbf{u}$ by \eqref{eq:opt_u} with fixed $\mathbf{\Phi}$ and $\omega$\;
        Update $\mathbf{\Phi}$ by solving \textbf{P2-2} with fixed $\mathbf{u}$ and $\omega$\;
        Update $\omega$ by \eqref{eq:opt_omega} with fixed $\mathbf{u}$ and $\mathbf{\Phi}$\;
    }
    \KwRet{$\mathbf{\Phi}$.}
\end{algorithm}

\subsection{WMMSE-based RCM Optimization for Passive RIS}

For the passive RIS, if the amplitudes of the reflecting coefficients are allowed to be smaller than $1$, the reflecting coefficients of the passive RIS can be directly obtained by the WMMSE-based algorithm in Section \ref{subsec:WMMSE_active}, while setting $\sigma_1^2=0$, $a_{\max}=1$ and dropping constraint \eqref{eq:p1b}. For the passive RIS with unit modulus reflecting coefficients, the WMMSE-based algorithm needs some modifications to tackle the non-convex unit modulus constraints on the reflecting coefficients, i.e., $|\phi_m|=1, m = 1, \cdots, M$. Specifically, the reflecting coefficient optimization step in each iteration of the WMMSE-based algorithm for passive RIS can be modified as follows. Let $\sigma_1^2=0$ and drop constraint \eqref{eq:p1b}, \textbf{P2-2} for the passive RIS becomes
\begin{subequations}
    \begin{align}
        \textbf{P2-2-P:} & \min_{\bar{\pmb{\phi}}} \bar{\epsilon}(\bar{\pmb{\phi}}) \\
        &\text{s.t. } |\bar{\pmb{\phi}}_{m}|= 1, \forall m = 1, \cdots, M,\\
        &\quad\  \bar{\pmb{\phi}}_{M+1}=1.
    \end{align}
\end{subequations}
To deal with the unit-modulus constraint, we can exploit the semidefinite relaxation (SDR) technique to transform \textbf{P2-2-P} into an SDP problem. Specifically, let $\bar{\bar{\pmb{\phi}}}=[\bar{\pmb{\phi}}^T, 1]^T$ and $\mathbf{V} = \bar{\bar{\pmb{\phi}}}\bar{\bar{\pmb{\phi}}}^H$, we have
\begin{subequations}
    \begin{align}
        \textbf{P2-2-P-SDP:} & \min_{\mathbf{V}} \Tr(\mathbf{A}_{\rm p}
        \mathbf{V}) \\
        &\text{s.t. } \mathbf{V}\succeq \mathbf{0},\\
        &\quad\  \mathbf{V}_{m, m}=1, m=1, \cdots, M+2,
    \end{align}
\end{subequations}
where $\mathbf{A}_{\rm p}=p_0\bar{\mathbf{u}}_0\bar{\mathbf{u}}_0^H+\sum_{k=1}^{K}\zeta_kp_k\bar{\mathbf{u}}_k\bar{\mathbf{u}}_k^H$, with
\begin{equation*}
    \bar{\mathbf{u}}_0^H = [\mathbf{u}^H\mathbf{A}_0, -1], 
    \bar{\mathbf{u}}_k^H = [\mathbf{u}^H\mathbf{A}_k, 0], \forall k = 1, \cdots, K.
\end{equation*}
By solving \textbf{P2-2-P-SDP} with the convex optimization tools such as CVX \cite{grant2014cvx}, we finally can recover the rank-$1$ solution with unit modulus with the Gaussian randomization method in \cite{jia2021intelligent}. Therefore, the WMMSE-based algorithm in Algorithm \ref{Alg:WMMSE} can also be used to optimize the RCM of passive RIS, while the step updating $\boldsymbol{\Phi}$ needs to be realized by solving \textbf{P2-2-P-SDP} instead.

\section{How much power budget is required to realize a given target detection probability?}\label{sec:activeVSpassive}

From Section \ref{subsec:consum_model}, we can see that the active RIS also requires power consumption for the signal amplification operations, in addition to the power consumed by the control and phase shift switch circuits of the REs, which is dependent on the number of REs. Therefore, in the RIS-assisted spectrum sensing systems, the performance gains brought by the active RIS and passive RIS both depend on the RIS's power budget. To fairly compare passive RIS and active RIS, in this section, we aim to figure out whether the active RIS is a more energy-efficient solution for achieving a given target detection probability. Hence, we investigate the minimum power budget required to realize a target detection probability for the active RIS and passive RIS. Particularly, we focus on a special case with neglected direct links for ease of analysis. Besides, we consider that $\mathbf{f}_k$ and $\mathbf{G}$ are LoS channels to derive closed-form expressions. In such a setup, the channel between the PU/interferers and RIS is denoted by 
\begin{equation*}
    \mathbf{f}_k=\sqrt{\beta_{f,k}}\mathbf{a}^{\text{h}}(\theta_k^{\text{AOA}}, \psi_k^{\text{AOA}})\otimes \mathbf{a}^{\text{v}}(\theta_k^{\text{AOA}}, \psi_k^{\text{AOA}}), \ k = 0, \cdots, K,
\end{equation*}
where $\beta_{f,k}$ denotes the pathloss, $\theta_k^{\text{AOA}}$ and $\psi_k^{\text{AOA}}$ are respectively the azimuth angle of arrival (AOA) and elevation AOA, $\mathbf{a}^{\text{h}}$ and $\mathbf{a}^{\text{v}}$ are the steering vectors defined as
\begin{align*}
    \mathbf{a}^{\text{h}}(\theta, \psi)&=[1, \cdots, e^{-j\frac{2\pi d}{\lambda}(M^{\text{h}}-1)\sin(\theta)\cos(\psi)}]^T, \\
    \mathbf{a}^{\text{v}}(\theta, \psi)&=[1, \cdots, e^{-j\frac{2\pi d}{\lambda}(M^{\text{v}}-1)\cos(\theta)\cos(\psi)}]^T,
\end{align*}
where $d$ denotes the element space of the RIS that is usually set as $\lambda/2$. The RIS-SU channel can be written as
\begin{equation}\label{eq:LoSG}
    \mathbf{G} = \sqrt{\beta_{G}}\mathbf{a}_G\mathbf{b}_G^H,
\end{equation}
where 
\begin{equation*}
    \mathbf{a}_G = [1, \cdots, e^{-j\frac{2\pi d}{\lambda}(N-1)\sin(\theta_G^{\text{AOA}})}],
\end{equation*}
\begin{equation*}
    \mathbf{b}_G = \mathbf{a}^{\text{h}}(\theta_G^{\text{AOD}}, \psi_G^{\text{AOD}})\otimes \mathbf{a}^{\text{v}}(\theta_G^{\text{AOD}}, \psi_G^{\text{AOD}}),
\end{equation*}
$\theta_G^{\text{AOD}}$ and $\psi_G^{\text{AOD}}$ are the azimuth angle of departure (AOD) and elevation AOD of $\mathbf{G}$ at the RIS, and $\theta_G^{\text{AOA}}$ is the AOA of $\mathbf{G}$ at the SU.

In the following, we will first show that, when the interferers are negligible, an optimal amplification factor can be observed, which mainly depends on the ratio of the power consumption to the input power of each active RE. When considering the impacts of the interferers, we show that the active RIS's reflecting coefficients can be configured with some heuristic receiver design principles for multi-antenna systems, e.g., matched-filter (MF), zero-forcing (ZF), and minimum mean square error (MMSE).

\subsection{With Negligible Interferers}\label{subsec:required_power_woPUEA}

For the case where the impacts of the interferers are negligible, \textbf{P1} is equivalent to
\begin{subequations}
    \begin{align}
        \textbf{P3:} & \max_{\mathbf{\Phi}}\mathbf{f}_0^H\mathbf{\Phi}^H\mathbf{G}^H(\sigma_1^2\mathbf{G}\mathbf{\Phi}\mathbf{\Phi}^H\mathbf{G}^H+ \sigma_2^2\mathbf{I})^{-1}\mathbf{G}\mathbf{\Phi}\mathbf{f}_0 \label{eq:p2a} \\
        & \text{s.t. } \eqref{eq:p1b}, \eqref{eq:p1c}.
    \end{align}
\end{subequations}
Using the identity $(\mathbf{I}+\mathbf{AB})^{-1}=\mathbf{I}-\mathbf{A}(\mathbf{I}+\mathbf{BA})^{-1}\mathbf{B}$,
\eqref{eq:p2a} can be rewritten as
\begin{equation}\label{eq:Pt1}
    \max_{\mathbf{\Phi}}\frac{1}{\sigma_1^2}\mathbf{f}_0^H\left[\mathbf{I}-\left(\mathbf{I}+\frac{\sigma_1^2}{\sigma_2^2}\mathbf{\Phi}^H\mathbf{G}^H\mathbf{G\Phi}\right)^{-1}\right]\mathbf{f}_0,
\end{equation}
By substituting \eqref{eq:LoSG} into \eqref{eq:Pt1}, we have
\begin{align*}
    \left(\mathbf{I}+\frac{\sigma_1^2}{\sigma_2^2}\mathbf{\Phi}^H\mathbf{G}^H\mathbf{G\Phi}\right)^{-1} &= \left(\mathbf{I}+\frac{N\beta_{G}\sigma_1^2}{\sigma_2^2}\mathbf{\Phi}^H\mathbf{b}_G\mathbf{b}_G^H\mathbf{\Phi}\right)^{-1}.
\end{align*}
Denoting $C_0\triangleq N\beta_{G}\sigma_1^2/\sigma_2^2$, using Sherman-Morrison formula, we have 
\begin{align*}
    \left(\mathbf{I}+C_0\mathbf{\Phi}^H\mathbf{b}_G\mathbf{b}_G^H\mathbf{\Phi}\right)^{-1}= \mathbf{I}-\frac{C_0\mathbf{\Phi}^H\mathbf{b}_G\mathbf{b}_G^H\mathbf{\Phi}}{1+C_0\mathbf{b}_G^H\mathbf{\Phi}\mathbf{\Phi}^H\mathbf{b}_G}.
\end{align*}
Therefore, \textbf{P3} can be equivalently rewritten as
\begin{subequations}
\begin{align}
    \textbf{P3-1:} & \max_{\mathbf{\Phi}} \frac{|\mathbf{b}_G^H\mathbf{\Phi}\mathbf{a}_f|^2}{1+C_0\mathbf{b}_G^H\mathbf{\Phi}\mathbf{\Phi}^H\mathbf{b}_G} \label{eq:p22a} \\
    & \text{s.t. } \eqref{eq:p1b}, \eqref{eq:p1c}.
\end{align}
\end{subequations}

Obviously, the optimal phase shifts of the reflecting coefficients are given by
\begin{equation*}
    \theta_{m} = {\rm arg}(\mathbf{b}_{G}(m))- {\rm arg}(\mathbf{a}_{f}(m)), \forall m=1,\cdots, M.
\end{equation*}
With the optimal phase shifts, we have 
\begin{equation*}
    \frac{|\mathbf{b}_G^H\mathbf{\Phi}\mathbf{a}_f|^2}{1+C_0\mathbf{b}_G^H\mathbf{\Phi}\mathbf{\Phi}^H\mathbf{b}_G}=\frac{(\sum_{m=1}^{M}a_m)^2}{1+C_0\sum_{m=1}^{M}a_m^2}
    \leq \frac{M^2a^2}{1+C_0Ma^2},
\end{equation*}
where the equality holds if and only if $a_1=\cdots=a_M=a$ according to Cauchy–Schwarz inequality. For a given power budget $P_{\rm ARIS}$, the maximum objective value of \eqref{eq:p22a} depends on not only $a$ but also $M$, which should satisfy
\begin{equation*}
    Ma^2\beta_{f,0}p_0+Ma^2\sigma_1^2+M(P_{\rm C}+P_{\rm DC})\leq P_{\rm ARIS}.
\end{equation*}
Suppose that the number of REs can be an arbitrary value, the optimal $M$ can be obtained only when $P_{\rm ARIS}$ is fully consumed by the $M$ REs. This can be easily verified via proof by contradiction as follows. Once the power consumption of the active RIS is lower than $P_{\rm ARIS}$, we can increase $M$ or $a$ to make \eqref{eq:p22a} larger. Therefore, denoting $A = a^2$, $C_1=P_{\rm C}+P_{\rm DC}$, and $C_2=\beta_{f,0}p_0+\sigma_1^2$, we have
\begin{equation}\label{eq:relationship_MA}
M = \frac{P_{\rm ARIS}}{C_1+C_2A},
\end{equation}
To maximize $M^2a^2/(1+C_0Ma^2)$ for a given power budget $P_{\rm ARIS}$, we obtain an optimization problem as 
\begin{subequations}
\begin{align}
    & \max_{A} \frac{AP_{\rm ARIS}^2}{(C_1+C_2A)(C_1+C_2A+C_0AP_{\rm ARIS})}\label{eq:p4-1a} \\
    & \text{s.t. } A\leq a_{\rm max}^2. \label{eq:p4-1b}
\end{align}
\end{subequations}
Dividing both the numerator and denominator of \eqref{eq:p4-1a} by $A$, \eqref{eq:p4-1a} can be transformed as 
\begin{equation*}
    \frac{P_{\rm ARIS}^2}{\frac{C_1^2}{A}+C_2(C_2+C_0P_{\rm ARIS})A+2C_1C_2+C_0C_1P_{\rm ARIS}}.
\end{equation*}
Regardless of constraint \eqref{eq:p4-1b}, we can obtain the optimal $A$ by utilizing the first-order optimality condition, and the unconstrained optimal $A$ is denoted by $A_0=C_1C_2^{-\frac{1}{2}}(C_2+C_0P_{\rm ARIS})^{-\frac{1}{2}}$. Under constraint \eqref{eq:p4-1b}, we have
\begin{equation*}
    A_{\rm opt}=\left\{
        \begin{array}{ccl}
            A_0,       &      &{\rm if}\  A_0< a_{\rm max}^2,\\
        a_{\rm max}^2,     &      & {\rm if}\ A_0\geq a_{\rm max}^2.
        \end{array} \right.
\end{equation*}
Eventually, the optimal $a$ is given by
\begin{equation}\label{eq:opt_a}
    a_{\rm opt} = \min \left\{a_{\rm max}, \sqrt{A_0}\right\}, 
\end{equation}
and the optimal $M$ is
\begin{equation}\label{eq:opt_M}
    M_{\rm opt} = \frac{P_{\rm ARIS}}{P_{\rm C}+P_{\rm DC}+a_{\rm opt}^2(\beta_{f,0}p_0+\sigma_1^2)}.
\end{equation}
Obviously, $a_{\rm opt}$ and $M_{\rm opt}$ depend on the relationship between $a_{\max}$ and $\sqrt{A_0}$, which means that the active RIS spectrum sensing performance can be improved if a larger $a_{\max}$ can be realized while satisfying $a_{\max}\leq \sqrt{A_0}$.

When $M_{\rm opt}$ is not an integer such that $M_{\rm opt}\in(M_{0}, M_0+1)$, the optimal integral number of active REs and the corresponding amplification factor, which are respectively denoted by $\bar{M}_{\rm opt}$ and $\bar{a}_{\rm opt}$, can be recovered as follows. By defining
\begin{equation*}
    \xi(M) = \sqrt{\frac{P_{\rm ARIS}-MC_1}{MC_2}},\  q(M, a)=\frac{M^2a^2}{1+C_0Ma^2},
\end{equation*}
when $a_{\rm opt} = a_{\rm max}$, we have $\bar{M}_{\rm opt} = M_0$, and $\bar{a}_{\rm opt}=a_{\max}$. When $a_{\rm opt} =\sqrt{A_0}$, we have $\bar{M}_{\rm opt}=M_0$, $\bar{a}_{\rm opt}=\min\{a_{\max}, \xi(M_0)\}$ if $q(M_0, \min\{a_{\max}, \xi(M_0)\})>q(M_0+1, \xi(M_0+1))$, and otherwise we have $\bar{M}_{\rm opt}=M_0+1$, $\bar{a}_{\rm opt}=\xi(M_0+1)$.

In the following, we analyze the required power budget to realize a target detection probability for the active RIS and passive RIS. In particular, we use $a_{\rm opt}$ and $M_{\rm opt}$ as the optimal configurations of the active RIS for simplicity. With $a_{\rm opt}$ and $M_{\rm opt}$, we denote the largest eigenvalue of the population covariance matrix under $\mathcal{H}_1$ by $\eta_{\rm ARIS}+1$ with
\begin{align}
    &\eta_{\rm ARIS}=p_0\mathbf{f}_0^H\mathbf{\Phi}^H\mathbf{G}^H(\sigma_1^2\mathbf{G}\mathbf{\Phi}\mathbf{\Phi}^H\mathbf{G}^H+ \sigma_2^2\mathbf{I})^{-1}\mathbf{G}\mathbf{\Phi}\mathbf{f}_0\nonumber\\
    &=\frac{a_{\rm opt}^2p_0M_{\rm opt}^2\beta_{f,0}\beta_{G}}{\sigma_2^2}\mathbf{a}_{G}^H(M_{\rm opt}\sigma_1^2a_{\rm opt}^2\beta_{G}\mathbf{a}_{G}\mathbf{a}_{G}^H+ \sigma_2^2\mathbf{I})^{-1}\mathbf{a}_{G}\nonumber\\
    &=\frac{a_{\rm opt}^2p_0M_{\rm opt}\beta_{f,0}}{\sigma_1^2a_{\rm opt}^2}\left[1-\left(1+\frac{M_{\rm opt}\sigma_1^2a_{\rm opt}^2\beta_{G}\mathbf{a}_{G}^H\mathbf{a}_{G}}{\sigma_2^2}\right)^{-1}\right]\nonumber\\
    &=\frac{a_{\rm opt}^2M_{\rm opt}^2N\beta_{f,0}\beta_{G}p_0}{M_{\rm opt}N\sigma_1^2a_{\rm opt}^2\beta_{G}+\sigma_2^2}\nonumber\\
    &=\frac{Na_{\rm opt}^2P_{\rm ARIS}^2\beta_{f,0}\beta_{G}p_0}{Na_{\rm opt}^2\beta_{G}\sigma_1^2P_{\rm ARIS}P_{\rm RE}+\sigma_2^2P_{\rm RE}^2},
\end{align}
where $P_{\rm RE}=P_{\rm C}+P_{\rm DC}+a_{\rm opt}^2(\beta_{f,0}p_0+\sigma_1^2)$ denotes the total power consumption of an RE with amplification factor $a_{\rm opt}$.

On the other hand, the power consumption model of the passive RIS is given by
\begin{equation}
    P_{\rm PRIS} = MP_{\rm C}.
\end{equation}
With the optimal phase shift design $\theta_{m} = {\rm arg}(\mathbf{b}_{G}(m))- {\rm arg}(\mathbf{a}_{f}(m))$, the largest eigenvalue of the population covariance matrix under $\mathcal{H}_1$ is $\eta_{\rm PRIS}+1$ with
\begin{equation}
    \eta_{\rm PRIS}=\frac{NM^2\beta_{f,0}\beta_{G}p_0}{\sigma_2^2}=\frac{NP_{\rm PRIS}^2\beta_{f,0}\beta_{G}p_0}{P_{\rm C}^2\sigma_2^2}.
\end{equation}

In the RIS-assisted spectrum sensing system, with a larger power budget, more REs can be supported for both passive and active RIS, and a higher amplification factor may also be achieved for the active RIS. Therefore, the largest eigenvalue of the population covariance matrix can be enhanced, leading to better spectrum sensing performance. Consequently, the required RIS power budget to achieve a target detection probability can be obtained with the bisection method. On the other hand, the detection probability can be evaluated by the distribution of the largest sample eigenvalue, whose relation to the largest population eigenvalue can be characterized by the spiked model from random matrix theory, which unfolds as the following lemma \cite{ge2022ris}.
\begin{lemma}\label{lemma:spiked_model}
    Let $\chi=N/T$ with large $N$ and $T$, denoting the largest population eigenvalue by $\eta_{\rm RIS}+1$, the distribution of the largest sample eigenvalue $\lambda_{\max}$ can be described as follows.
    \begin{itemize}
        \item When $\eta_{\rm RIS}<\sqrt{\chi}$,
        \begin{equation*}
            N^{\frac{2}{3}}\frac{\lambda_{\max}-(1+\sqrt{\chi})^2}{(1+\sqrt{\chi})^{\frac{4}{3}}\sqrt{\chi}}\xrightarrow[N,T\to\infty]{\mathcal{D}} F_2,
        \end{equation*}
        where $F_2$ is the Tracy-Widom distribution of order $2$.
        \item When $\eta_{\rm RIS}>\sqrt{\chi}$, $\lambda_{\max}$ satisfies a Gaussian distribution as $\lambda_{\max}\sim \mathcal{N}\left(\mu_{a}(\eta_{\rm RIS}), v_{a}(\eta_{\rm RIS})\right)$,
        where
        \begin{align*}
            \mu_{a}(\eta_{\rm RIS}) &= \eta_{\rm RIS}+1+\chi+\frac{\chi}{\eta_{\rm RIS}},\\
            v_{a}(\eta_{\rm RIS}) &= \frac{(\eta_{\rm RIS}+1)^2}{n}\left(1-\frac{\chi}{\eta_{\rm RIS}}\right).
        \end{align*}
    \end{itemize}
\end{lemma}
It should be noted that we have $\eta_{\rm RIS}=0$ for hypothesis $\mathcal{H}_0$, which can be regarded as a special case of $\eta_{\rm RIS}<\sqrt{\chi}$. From Lemma 2, the distribution of the test statistic under hypothesis $\mathcal{H}_1$ is the same as hypothesis $\mathcal{H}_0$. In such cases where $\eta_{\rm RIS}<\sqrt{\chi}$, the presence of the PU almost can not be identified, leading to a detection probability that equals the false alarm probability. In spectrum sensing problems, we aim to achieve a high detection probability to protect the PU while keeping a low false alarm probability. Therefore, to achieve a high detection probability, e.g., a detection probability of $0.9$ is required in IEEE 802.22 standard \cite{stevenson2009ieee}, we have to realize an $\eta_{\rm RIS}$ larger than $\sqrt{\chi}$, and the minimum $\eta_{\rm RIS}$ for achieving a target probability of $\bar{P}_{d}$ should satisfy
\begin{equation}\label{eq:min_eta}
    Q\left(\frac{\gamma_{\rm th}-\mu_{a}(\eta_{\rm RIS})}{\sqrt{v_{a}(\eta_{\rm RIS})}}\right)=\bar{P}_{d}.
\end{equation}
Let $\eta_0$ denote the solution to \eqref{eq:min_eta}, the minimum power budget for realizing $\eta_0$ is exactly that for achieving a target probability of $\bar{P}_{d}$, which can be obtained via the bisection method due to the fact that $\eta_{\rm RIS}$ is positively related with the power budget for both active and passive RISs.

\subsection{With Non-negligible Interferers}\label{subsec:required_power_wPUEA}

From the above analysis, the required power budget of the active RIS and that of the passive RIS can be obtained under the simplified case with negligible interferers. To figure out the impacts of the interferers on the required power budget, we consider the existence of the interferers in the following analysis. 
Let $\mathbf{D}=(\sigma_1^2\mathbf{I}_M+\sum_{k=1}^{K}\zeta_kp_k\mathbf{f}_k\mathbf{f}_k^H)/\sigma_2^2$, \textbf{P1} reduces to
\begin{subequations}
    \begin{align}
        \textbf{P4:} & \max_{\mathbf{\Phi}}\mathbf{f}_0^H\mathbf{\Phi}^H\mathbf{G}^H\left(\mathbf{I}_N +\mathbf{G}\mathbf{\Phi}\mathbf{D}\mathbf{\Phi}^H\mathbf{G}^H\right)^{-1}\mathbf{G}\mathbf{\Phi}\mathbf{f}_0 \label{eq:p4a} \\
        & \text{s.t. } \eqref{eq:p1b}, \eqref{eq:p1c}.
    \end{align} 
\end{subequations}
Using the Woodbury identity \cite{golub2013matrix}, namely, 
\begin{align*}
    &(\mathbf{A}+\mathbf{C}\mathbf{B}\mathbf{C}^H)^{-1}\\
    &\qquad=\mathbf{A}^{-1}-\mathbf{A}^{-1}\mathbf{C}(\mathbf{B}^{-1}+\mathbf{C}^H\mathbf{A}^{-1}\mathbf{C})^{-1}\mathbf{C}^H\mathbf{A}^{-1},
\end{align*}
we have
\begin{align*}
    &\left(\mathbf{I}_N+\mathbf{G}\mathbf{\Phi}\mathbf{D}\mathbf{\Phi}^H\mathbf{G}^H\right)^{-1} \\
    &\qquad=\mathbf{I}_N-\mathbf{G}\mathbf{\Phi}(\mathbf{D}^{-1}+\mathbf{\Phi}^H\mathbf{G}^H\mathbf{G}\mathbf{\Phi})^{-1}\mathbf{\Phi}^H\mathbf{G}^H.
\end{align*}
Therefore, \eqref{eq:p4a} can be expressed as
\begin{equation*}
    \mathbf{f}_0^H\mathbf{\Phi}^H\mathbf{G}^H\!\!\left[\mathbf{I}_N\!-\!\mathbf{G}\mathbf{\Phi}(\mathbf{D}^{-1}\!+\!\mathbf{\Phi}^H\mathbf{G}^H\mathbf{G}\mathbf{\Phi})^{-1}\mathbf{\Phi}^H\mathbf{G}^H\right]\mathbf{G}\mathbf{\Phi}\mathbf{f}_0.
\end{equation*}
Let $\mathbf{E}\triangleq \mathbf{\Phi}^H\mathbf{G}^H\mathbf{G}\mathbf{\Phi}$, \eqref{eq:p4a} can be finally rewritten as
\begin{equation}\label{eq:transform1}
    \mathbf{f}_0^H\left(\mathbf{E}-\mathbf{E}(\mathbf{D}^{-1}+\mathbf{E})^{-1}\mathbf{E}\right)\mathbf{f}_0.
\end{equation}
With the identity $\mathbf{A}-\mathbf{A}(\mathbf{A}+\mathbf{B})^{-1}\mathbf{A}=\mathbf{B}-\mathbf{B}(\mathbf{A}+\mathbf{B})^{-1}\mathbf{B}$, \eqref{eq:transform1} can be equivalently transformed as
\begin{equation}\label{eq:trans1}
    \mathbf{f}_0^H\left(\mathbf{D}^{-1}-\mathbf{D}^{-1}(\mathbf{D}^{-1}+\mathbf{E})^{-1}\mathbf{D}^{-1}\right)\mathbf{f}_0.
\end{equation}
Noting that $\mathbf{G}$ is a rank-$1$ matrix under assumption \eqref{eq:LoSG},
\begin{align*}
    \mathbf{E} &= \mathbf{\Phi}^H\mathbf{G}^H\mathbf{G}\mathbf{\Phi} = \beta_G\mathbf{\Phi}^H\mathbf{b}_G\mathbf{a}_G^H\mathbf{a}_G\mathbf{b}_G^H\mathbf{\Phi}\\
    & = N\beta_G\mathbf{\Phi}^H\mathbf{b}_G\mathbf{b}_G^H\mathbf{\Phi}
\end{align*}
is also a rank-$1$ matrix. Using Sherman-Morrison formula, we can derive
\begin{align}
    (\mathbf{D}^{-1}+\mathbf{E})^{-1} &= \left(\mathbf{D}^{-1}+N\beta_G\mathbf{\Phi}^H\mathbf{b}_G\mathbf{b}_G^H\mathbf{\Phi}\right)^{-1}\nonumber\\
    &= \mathbf{D}-\frac{\mathbf{D}N\beta_G\mathbf{\Phi}^H\mathbf{b}_G\mathbf{b}_G^H\mathbf{\Phi}\mathbf{D}}{1+N\beta_G\mathbf{b}_G^H\mathbf{\Phi}\mathbf{D}\mathbf{\Phi}^H\mathbf{b}_G}.\label{eq:SM_trans}
\end{align}
By substituting \eqref{eq:SM_trans} into \eqref{eq:trans1}, \eqref{eq:trans1} becomes 
\begin{equation*}
    \frac{N\beta_G\mathbf{f}_0^H\mathbf{\Phi}^H\mathbf{b}_G\mathbf{b}_G^H\mathbf{\Phi}\mathbf{f}_0}{1+N\beta_G\mathbf{b}_G^H\mathbf{\Phi}\mathbf{D}\mathbf{\Phi}^H\mathbf{b}_G}.
\end{equation*}
Therefore, \textbf{P4} can be equivalently transformed as
\begin{subequations}
    \begin{align}
        \textbf{P4-1:} & \max_{\mathbf{\Phi}}\frac{\mathbf{b}_G^H\mathbf{\Phi}\mathbf{f}_0\mathbf{f}_0^H\mathbf{\Phi}^H\mathbf{b}_G}{1+N\beta_G\mathbf{b}_G^H\mathbf{\Phi}\mathbf{D}\mathbf{\Phi}^H\mathbf{b}_G} \\
        & \text{s.t. } \eqref{eq:p1b}, \eqref{eq:p1c}.
    \end{align}
\end{subequations}
Let $\boldsymbol{\phi}$ be a vector consisting of the diagonal elements of $\mathbf{\Phi}^H$, and defining a diagonal matrix $\mathbf{B} = {\rm diag}(\mathbf{b}_G)$, \textbf{P4-1} can be rewritten as
\begin{subequations}
    \begin{align}
        \textbf{P4-2:} & \max_{\mathbf{\Phi}}\frac{\boldsymbol{\phi}^H\mathbf{B}^H\mathbf{f}_0\mathbf{f}_0^H\mathbf{B}\boldsymbol{\phi}}{1+N\beta_G\boldsymbol{\phi}^H\mathbf{B}^H\mathbf{D}\mathbf{B}\boldsymbol{\phi}} \label{eq:p42a}\\
        & \text{s.t. } |\phi_m|\leq a_{\max}, \forall m = 1,\cdots, M, \label{eq:p42b}\\
        &\quad\  \boldsymbol{\phi}^H\boldsymbol{\phi} \leq \bar{P}_{\rm out}/\bar{P}_{\rm in} \label{eq:p42c},
    \end{align}    
\end{subequations} 
where constraint \eqref{eq:p42c} is an equivalent transformation of \eqref{eq:p1b}, and $\bar{P}_{\rm in}=\sum_{k=0}^{K}\zeta_k\beta_{f,k}p_k+\sigma_1^2$.

In particular, \eqref{eq:p42a} can be transformed into a generalized Rayleigh quotient form as
\begin{equation}\label{eq:generalized_RQ}
    \frac{\boldsymbol{\phi}^H\mathbf{B}^H\mathbf{f}_0\mathbf{f}_0^H\mathbf{B}\boldsymbol{\phi}}{\boldsymbol{\phi}^H(\frac{1}{\rho}\mathbf{I}+N\beta_G\mathbf{B}^H\mathbf{D}\mathbf{B})\boldsymbol{\phi}},
\end{equation}
where $\rho \triangleq \boldsymbol{\phi}^H\boldsymbol{\phi}$. Regardless of constraints \eqref{eq:p42b}, we can obtain a closed-form solution to maximize \eqref{eq:generalized_RQ} by letting $\rho=\bar{P}_{\rm out}/\bar{P}_{\rm in}$, and the solution can be expressed as
\begin{equation}\label{eq:phi_unconstrained}
    \boldsymbol{\phi}_{\rm MMSE} = \left(\frac{1}{\rho}\mathbf{I}+N\beta_G\mathbf{B}^H\mathbf{D}\mathbf{B}\right)^{-1}\mathbf{B}^H\mathbf{f}_0,
\end{equation}
which is also known as the MMSE-based solution. However, due to the per-element amplification factor constraint (PEAFC) \eqref{eq:p42b}, we may have to resort to the WMMSE algorithm to obtain a optimal solution, which makes it intractable to analyze the required power budget to achieve a given target detection probability. Alternatively, we relax the PEAFC \eqref{eq:p42b} as a constraint of $\rho$, namely, $\rho\leq Ma_{\max}^2$, for simplicity. As such, the optimal $\rho$ for the MMSE-based solution can be obtained by
\begin{equation*}
    \rho_{1} = \min\{\bar{P}_{\rm out}/\bar{P}_{\rm in}, Ma_{\max}^2\}.
\end{equation*}
By substituting \eqref{eq:phi_unconstrained} into \eqref{eq:p42a}, the maximized objective of \textbf{P4-2} is given by
\begin{equation*}
    \mathbf{f}_0^H\mathbf{B}\left(\frac{1}{\rho_{1}}\mathbf{I}+N\beta_G\mathbf{B}^H\mathbf{D}\mathbf{B}\right)^{-1}\mathbf{B}^H\mathbf{f}_0,
\end{equation*}
and $\eta_{\rm ARIS}$ with the MMSE-based solution is therefore
\begin{equation}\label{eq:eta_mmse}
    \eta_{\rm ARIS, MMSE}\! = \!\frac{N\beta_Gp_0}{\sigma_2^2}\mathbf{f}_0^H\mathbf{B}\left(\frac{1}{\rho_{1}}\mathbf{I}\!+\!N\beta_G\mathbf{B}^H\mathbf{D}\mathbf{B}\right)^{-1}\mathbf{B}^H\mathbf{f}_0.
\end{equation}
It is worth noting that the MMSE-based solution may violate the PEAFC, and $\eta_{\rm ARIS, MMSE}$ may can not be achieved. Therefore, $\eta_{\rm ARIS, MMSE}$ actually provides an upper bound of $\eta_{\rm ARIS}$ for all available solutions to \textbf{P4-2}.

In addition to the MMSE-based solution providing the upper bound performance, we can also derive several alternative solutions \textbf{P4-2}. Note that \textbf{P4-2} is similar to the receiver design problems for conventional multi-antenna systems, $\boldsymbol{\phi}$ can be alternatively designed with the well-known ZF and MF principles, where \eqref{eq:p42b} and \eqref{eq:p42c} can be satisfied by straightforwardly scaling the obtained $\boldsymbol{\phi}$. In particular, the ZF-based and MF-based solutions can be obtained as follows. Denoting $\mathbf{Q} = [\mathbf{q}_0, \cdots, \mathbf{q}_K]$, where $\mathbf{q}_k \triangleq \mathbf{B}^H\mathbf{f}_k$, let $\mathbf{w}$ be first column of $\mathbf{Q}(\mathbf{Q}^H\mathbf{Q})^{-1}$, the ZF-based solution satisfying \eqref{eq:p42b} and \eqref{eq:p42c} is given by 
\begin{equation}\label{eq:phi_zf}
    \boldsymbol{\phi}_{\rm ZF} = \sqrt{\rho_2}\frac{\mathbf{w}}{\|\mathbf{w}\|},
\end{equation}
where 
\begin{equation}\label{eq:rho_2}
    \rho_2 = \min\left\{\frac{\bar{P}_{\rm out}}{\bar{P}_{\rm in}}, \frac{a_{\max}^2\|\mathbf{w}\|^2}{\|\mathbf{w}\|_{\infty}^2}\right\}.
\end{equation}
It should be noted that $M\geq K+1$ is necessary in the ZF-based method to ensure that $\mathbf{Q}^H\mathbf{Q}$ is invertible, where $\eta_{\rm ARIS}$ can be written as  
\begin{align}
    \eta_{\rm ARIS, ZF} &= \frac{N\beta_Gp_0}{\left(\frac{\sigma_2^2}{\rho_2}+N\beta_G\sigma_1^2\right)[(\mathbf{Q}^H\mathbf{Q})^{-1}]_{1,1}} \nonumber\\
    &\hspace{-1em}=\frac{N\beta_G\rho_2p_0}{\sigma_2^2+N\beta_G\rho_2\sigma_1^2}\mathbf{q}_0^H(\mathbf{I}-\bar{\mathbf{Q}}(\bar{\mathbf{Q}}^H\bar{\mathbf{Q}})^{-1}\bar{\mathbf{Q}}^H)\mathbf{q}_0, \label{eq:eta_zf}
\end{align}
where $\bar{\mathbf{Q}} = [\mathbf{q}_1, \cdots, \mathbf{q}_K]$.

The idea behind the MF-based solution is only enhancing the signals from the interested PU. Therefore, the MF-based solution is exactly the same with the optimal solution under the case with negligible interferers, which can be expressed as
\begin{equation}\label{eq:phi_mf}
    \boldsymbol{\phi}_{\rm MF} = a\mathbf{B}^H\mathbf{a}_f,
\end{equation}
where
\begin{equation*}
    a = \min\left\{a_{\max}, \sqrt{\frac{\bar{P}_{\rm out}}{M\bar{P}_{\rm in}}}\right\}. 
\end{equation*}
Therefore, $\eta_{\rm ARIS}$ with the MF-based solution is
\begin{align}\label{eq:eta_mf}
    \eta_{\rm ARIS, MF} = \frac{NM^2a^2\beta_{f,0}\beta_Gp_0}{(1+Na^2\beta_G\mathbf{a}_f^H\mathbf{D}\mathbf{a}_f)\sigma_2^2}.
\end{align}

In the case with non-negligible interferers, for all the above solutions, $\eta_{\rm ARIS}$ can be optimized by increasing $M$ or upscaling $\boldsymbol{\phi}$ while satisfying constraints \eqref{eq:p42b} and \eqref{eq:p42c}. Hence, a tradeoff between $M$ and $\boldsymbol{\phi}$ still exists. However, it is difficult to straightforwardly figure out the optimal $M$ and $\boldsymbol{\phi}$ as in the case with negligible interferers, because $M$ is not explicitly expressed in \eqref{eq:eta_mmse}, \eqref{eq:eta_zf}, and \eqref{eq:eta_mf}. To find the optimal tradeoff realizing the largest $\eta_{\rm ARIS}$ for a given power budget, one straightforward way is exhausting all the available integral $M$'s. Particularly, the largest value among the available $M$'s is given by
\begin{equation}\label{eq:Mmax}
    M_{\max} \!=\! \left\lfloor \frac{P_{\rm ARIS}}{P_{\rm C}+P_{\rm DC}}\right\rfloor.
\end{equation}
As such, we can obtain the optimal combination of $\mathbf{\Phi}$ and $M$ to achieve the largest $\eta_{\rm ARIS}$ for a given RIS power budget. Then, following the conclusions drawn from Lemma \ref{lemma:spiked_model}, we can finally obtain the required power budget to achieve a target detection probability of $\bar{P}_{d}$, which can be summarized in Algorithm \ref{Alg:required_powerbuget}.

\begin{algorithm}[tbp]
    \caption{The bisection method to obtain the required power budget for achieving $\bar{P}_{d}$\label{Alg:required_powerbuget}}
    Obtain $\eta_0$ by solving \eqref{eq:min_eta}\;
    Initialize a large power budget $P_{\rm high}$, $P_{\rm low}=0$, and a stop condition $\epsilon_{\rm stop}$\;
    \While {$|P_{\rm high}-P_{\rm low}|>\epsilon_{\rm stop}$}{
        Update $P_{\rm ARIS} = (P_{\rm high}+P_{\rm low})/2$\;
        Calculate  $M_{\max}$ with \eqref{eq:Mmax}\;
        \For {$i = 1:M_{\max}$}{
            $M = i$\;
            Calculate $\eta_{i}$ with one of \eqref{eq:eta_mmse}, \eqref{eq:eta_zf}, and \eqref{eq:eta_mf}, depending on the adopted method to obtain $\boldsymbol{\Phi}$\;
        }
        $\eta^*=\max_i\{\eta_{i}\}$\;
        \eIf {$\eta^*>\eta_{0}$}{
            $P_{\rm high}=P_{\rm ARIS}$\;
        } {
            $P_{\rm low}=P_{\rm ARIS}$\;
        }
    }
    \KwRet{$P_{\rm ARIS}$.}
\end{algorithm}

\section{Simulation Results}\label{sec:sim}

In this section, we provide extensive simulations to demonstrate the performance of the proposed active RIS-assisted spectrum sensing system. Specifically, the number of antennas at the SU is set as $N=64$. The locations of the PU, active RIS, and SU are described with a two-dimensional Cartesian coordinate and respectively given by ($0$, $0$), ($100$m, $50$m), and ($500$m, $0$). Besides, we assume that $K=5$ interferers exist in a circular region with a minimum radius of $50$m and a maximum radius of $60$m, centered around the RIS. The pathloss of the PU/interferer-SU channel, the PU/interferer-RIS channel, and the RIS-SU channel are given by \cite{rappaport1996wireless}
\begin{equation*}
    \beta_{d,k} = \frac{\lambda^2}{(4\pi)^2d_{d, k}^{\alpha_1}},\ 
    \beta_{f,k} = \frac{\lambda^2}{(4\pi)^2d_{f, k}^{\alpha_2}},\ 
    \beta_G = \frac{\lambda^2}{(4\pi)^2d_G^{\alpha_3}},\ 
\end{equation*}
where $\lambda$ is the carrier wavelength, $d_{d, k}$, $d_{f, k}$, and $d_G$ denote the distances of the corresponding channels, $\alpha_1$, $\alpha_2$, and $\alpha_3$ are the pathloss exponents. In the simulations, we set $\lambda=0.12$m, corresponding to a carrier frequency of $2.5$GHz. The pathloss exponents are set with $\alpha_1=4$, $\alpha_2=2$, and $\alpha_3=2$. The transmitting power of the PU and the interferers is set as $p_0=p_1=\cdots=p_K=30$ dBm, and the power of the thermal noise at the active RIS and that of the AWGN at the SU are set as $\sigma_1^2=\sigma_2^2=-80$dBm. For the power consumption model of the RIS, the power consumption of the control and phase shift switch circuits for each RE is set as $P_{\rm C} = -10$dBm, and the direct current power consumption at each active RE is set as $P_{\rm DC} = -5$dBm \cite{long2021active}.

\begin{figure}[tbp]
    \begin{center}
        \epsfxsize=0.9\linewidth
        \epsffile{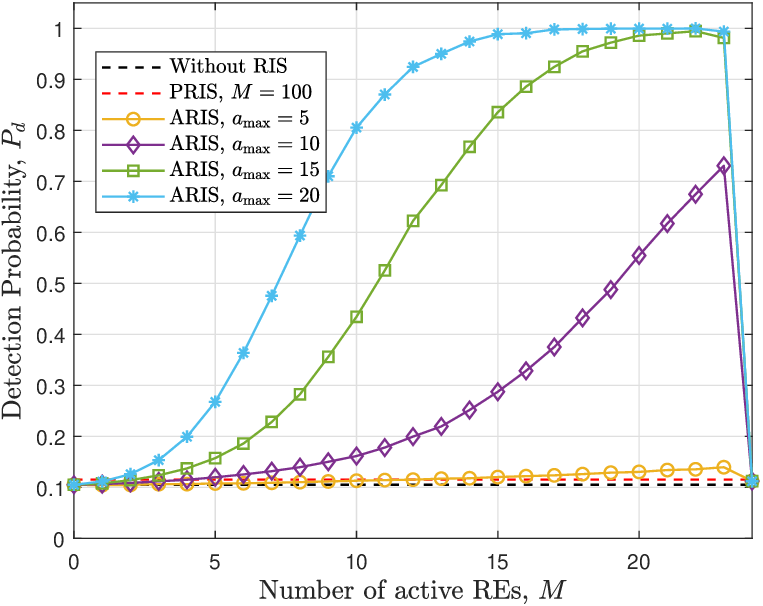}
        \caption{The performance gains of the active RIS and passive RIS on the detection probability.}\label{fig:perf_cmp}
    \end{center}
\end{figure}

Firstly, we investigate the performance gains of the active RIS and passive RIS on the detection probability, and the detection threshold is calculated by \eqref{eq:detection_th} with the maximum false alarm probability set as $\alpha=0.1$. The number of signal samples is set as $T=100N=6400$ such that $c=n/T=0.01$. Besides, the active probabilities of the interferers are set as $\zeta_{k}=1$ for arbitrary $k$. Fig. \ref{fig:perf_cmp} compares the detection probability improvements achieved by the active RIS and the passive RIS for a given power budget of $P_{\rm RIS}=10$dBm, which can support up to $100$ passive REs. The results are obtained by averaging $500$ random channel realizations, while the small-scale channel fading is characterized by Rayleigh fading. For the conventional spectrum sensing system without RIS, such a number of signal samples is far from enough for the SU to identify the presence of the primary signals, leading to a detection probability almost equal to the false alarm probability. The detection probability can be improved by introducing the passive RIS, but the improvement is minor because the given power budget is insufficient for the passive RIS. For the active RIS, we investigate the detection probability for different numbers of active REs, and the achieved detection probability increases first and then decreases as $M$ grows to the maximum $M$. In particular, the decreasing detection probability from $M=23$ to $M=24$ is because the remaining power budget is insufficient to support sufficiently large amplification factors, i.e., constraint \eqref{eq:p1b}. An optimal number of active REs achieving the best sensing performance can be observed for all $a_{\max}$. Moreover, we can see that the maximum detection probability achieved by the active RIS increases as $a_{\max}$ becomes larger. This observation is consistent with the analysis in Section \ref{sec:activeVSpassive}, i.e., when $a_{\max}$ becomes larger while being kept below a certain value, say $\sqrt{A_0}$, the superiority of the active RIS will be more significant.

\begin{figure}[tbp]
    \begin{center}
        \epsfxsize=0.9\linewidth
        \epsffile{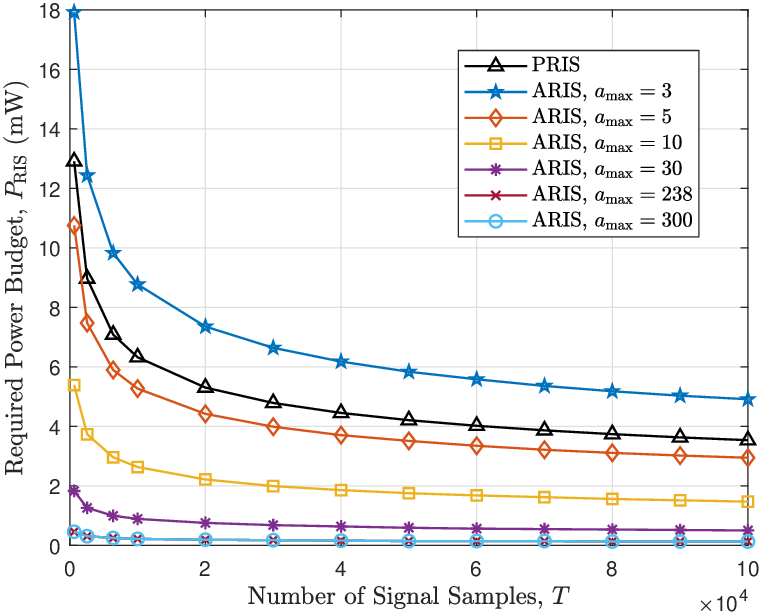}
        \caption{The required power budget to realize a target detection probability of $0.9$ for different $T$'s under the case with negligible interferers.}\label{fig:RBD_Ts_woPUEA}
    \end{center}
\end{figure}

In Fig. \ref{fig:RBD_Ts_woPUEA}, we investigate the existence of the optimal amplification factor under the special case without direct links and the existence of interferers. Particularly, we show the required power budget to realize a target detection probability of $0.9$ for different amounts of sensing signal samples, and the active RIS and passive RIS are compared for the LoS RIS-related channels. It can be observed that the required power budget of both active RIS and passive RIS decreases when the number of sensing signal samples increases. In other words, the required sensing time can be reduced by increasing the power budget of the RIS, thus the spectrum efficiency can also be improved. Besides, for a small $a_{\max}$ (e.g., $a_{\max}=3$), the active RIS requires more power budget as compared to the passive RIS. When $a_{\max}\geq 5$, the active RIS requires less power budget than the passive RIS, and the required power budget can be reduced by increasing $a_{\max}$, which means that a larger $a_{\max}$ enables more power-efficient solutions. The explanation for this phenomenon is as follows. Since the active RE consumes more power to amplify the incident signals, the passive RIS can employ much more passive REs for the same power budget. As a result, as long as the performance gain brought by the larger $a_{\max}$ cannot compensate for the performance loss caused by the reduction of the number of REs, the passive RIS will outperform the active RIS in terms of the required power budget for realizing the same target detection probability. Moreover, in the simulated scenario, we have $\sqrt{A_0}\approx \sqrt{C_1/C_2}=\sqrt{(P_{\rm C}+P_{\rm DC})/(\beta_{f,0}p_0+\sigma_1^2)}\approx 238$, it can also be observed that the required power budget can not be further reduced by increasing $a_{\max}$ when $a_{\max}\geq 238$. Hence, without considering the physical limitation of $a_{\max}$, the required power can be minimized by letting $a=\sqrt{A_0}$ in such cases.

\begin{figure}[tbp]
    \begin{center}
        \epsfxsize=0.9\linewidth
        \epsffile{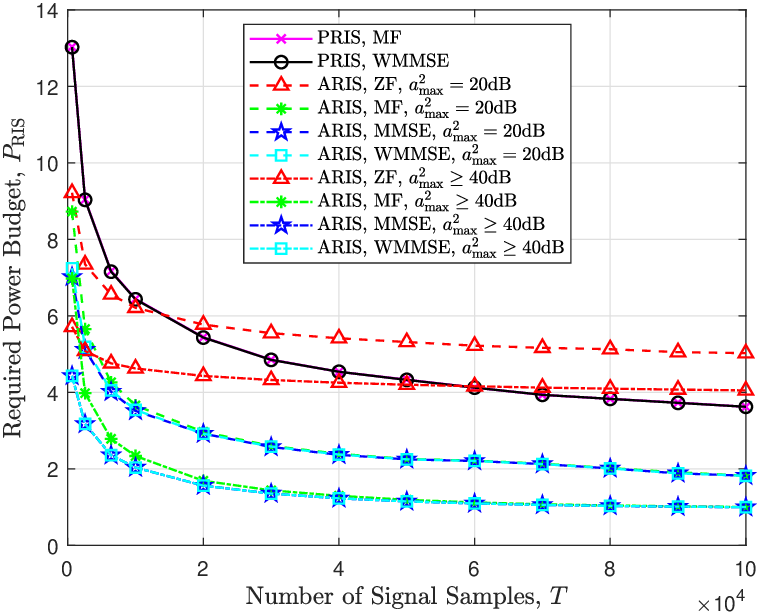}
        \caption{The required power budget to realize a target detection probability for different $T$'s under the case with non-negligible interferers.}\label{fig:RBD_Ts}
    \end{center}
\end{figure}

We also study the impacts of the interferers on the required RIS power budget in such a special case. The results under the scenario with $K=5$ interferers are depicted in Fig. \ref{fig:RBD_Ts}. In comparison with Fig. \ref{fig:RBD_Ts_woPUEA}, one straightforward observation is that, for both passive and active RISs, more power budget is required to realize $P_{d}=0.9$ while the interferers exist. Besides, in regard to the heuristic configuration methods (MF, ZF, and MMSE) for the active RIS, the required power budget can also be reduced by increasing $a_{\max}$. Moreover, for all the configuration methods, there exists an optimal $a_{\max}$, over which increasing $a_{\max}$ cannot further reduce the required power budget. The optimal $a_{\max}$ of the case with interferers is also smaller than that of the case without interferers. For the passive RIS, the performance of the MF-based RCM configuration can approach that of the WMMSE-based solution, demonstrating the near-optimality of the MF-based configuration method in this simulated scenario. In regard to the active RIS, the MF-based solution can not perform as well as the WMMSE-based solution when $T$ is small, but it can also approach the MMSE-based solution in the large $T$ region. Moreover, we can see that the required power budget of the WMMSE-based solution is almost the same as the MMSE-based lower bound for all $a_{\max}$, and the near-optimality of the WMMSE-based algorithm is validated. In addition, the ZF-based method aims to null the interference from the interferers no matter how strong the interference is, and $M\geq K+1$ is necessary to perform interference nulling, which means $(K+1)(P_{\rm C}+P_{\rm DC})$ is a lower bound of the zero-forcing method. Therefore, in the large $T$ region where the target $\eta$ is rather small, we can observe a floor of the required power budget of the ZF-based method, and it is much larger than the other active RIS configuration methods and the passive RIS.

\begin{figure}[tbp]
    \begin{center}
        \epsfxsize=0.9\linewidth
        \epsffile{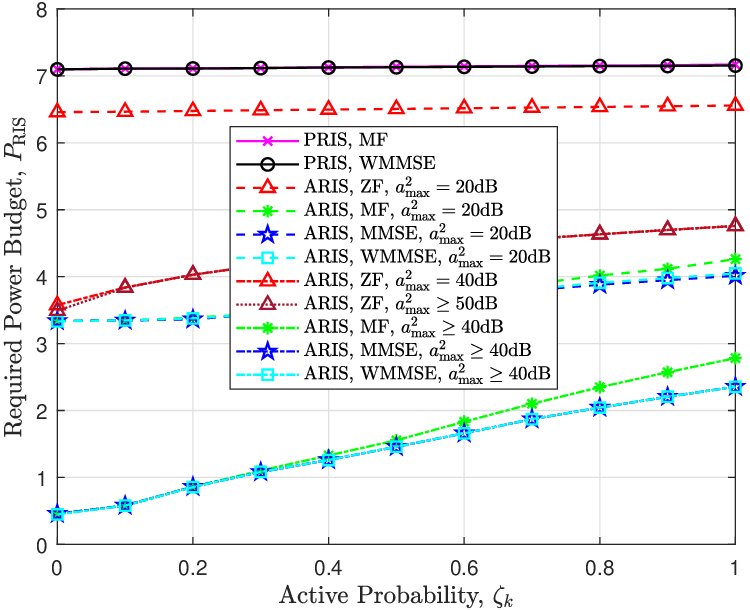}
        \caption{The required power budget to realize a target detection probability of $P_{d}=0.9$ for different active probabilities of the interferers.}\label{fig:RBD_active_prob}
    \end{center}
\end{figure}

Then, we delve into the impacts of the interferers' behaviors on the required power budget to achieve $P_d=0.9$. In Fig. \ref{fig:RBD_active_prob}, the required power budget with respect to the active probability of the interferers is depicted. When $\zeta_{k}=0$, the scenario with interferers reduces to the scenario without interferers. In such a case, the MF- and MMSE-based methods are exactly equivalent, and the required power budgets of these two methods are the same. As $\zeta_{k}$ increases, the required power budgets also increases, and the required power budget of the MF-based method starts to grow higher than the MMSE-based lower bound, which the WMMSE-based algorithm can still approach. Moreover, for the considered $a_{\max}^2$, namely, $20$dB and $40$dB, the active RIS configured by the MF-based and MMSE-based methods can outperform the passive RIS. As for the ZF-based method, when $a_{\max}^2=20$dB, the required power budget remains almost unchanged as $\zeta_{k}$ increases. In contrast, when $a_{\max}^2=40$dB or even larger than $50$dB, the required power budget of the ZF-based method also increases as $\zeta_{k}$ becomes larger. This phenomenon is quite counterintuitive because the performance of the ZF-based solution is independent of the interference strength in conventional multi-antenna systems. In the considered active RIS-assisted spectrum sensing system, this observation can be explained as follows. From \eqref{eq:rho_2} and \eqref{eq:eta_zf}, the achieved $\eta_{\rm ARIS, ZF}$ of the ZF-based solution mainly depends on $\rho_2$, which is actually determined by $a_{\max}$ and $\bar{P}_{\rm out}/\bar{P}_{\rm in}$. It should be noted that $\bar{P}_{\rm in}$ is the input signal power of the active RIS, which is related to $\zeta_{k}$. For a small $a_{\max}$ such that $a_{\max}^2\|\mathbf{w}\|^2/\|\mathbf{w}\|_{\infty}^2<\bar{P}_{\rm out}/\bar{P}_{\rm in}$, the required power budget is therefore independent of $\zeta_{k}$. For a large $a_{\max}$ leading to $a_{\max}^2\|\mathbf{w}\|^2/\|\mathbf{w}\|_{\infty}^2<\bar{P}_{\rm out}/\bar{P}_{\rm in}$ and $\rho_2=\bar{P}_{\rm out}/\bar{P}_{\rm in}$, with increase of $\zeta_{k}$, $\bar{P}_{\rm in}$ becomes larger, and $\bar{P}_{\rm out}$ has to become larger to achieve the same sensing performance, which finally results in an increasing power budget of the ZF-based solution. Besides, in this case, for each active RIS configuration method, the active RIS with a large enough $a_{\max}$ can outperform the passive RIS.

\begin{figure}[tbp]
    \begin{center}
        \epsfxsize=0.9\linewidth
        \epsffile{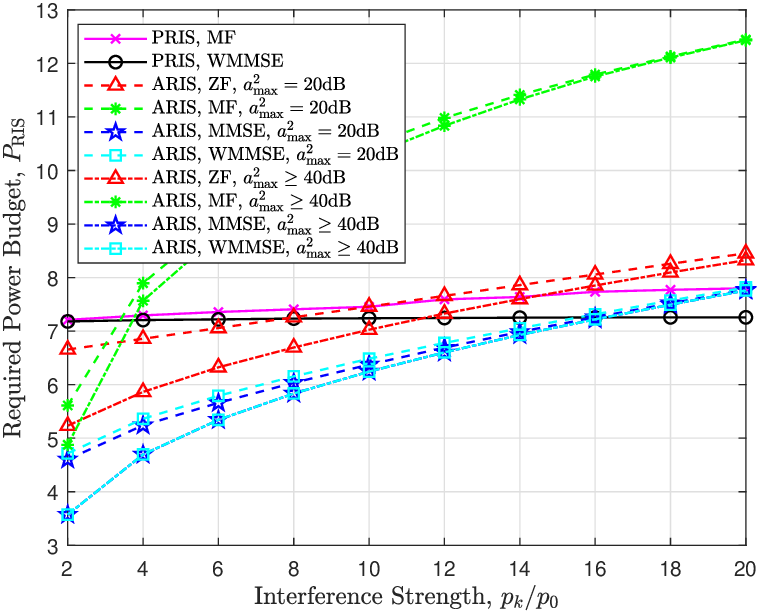}
        \caption{The required power budget to realize a target detection probability of $P_{d}=0.9$ for different transmit power of the interferers.}\label{fig:RBD_PUEA_Ps}
    \end{center}
\end{figure}

Furthermore, we investigate the scenario where the interference from the interferers is stronger than the interested primary signals. Without loss of generality, we assume that the interferers are always active, and they have the same transmit power, i.e., $\zeta_{1}=\cdots=\zeta_{K}=1$ and $p_{1}=\cdots=p_{K}$. Fig. \ref{fig:RBD_PUEA_Ps} depicts the relationship between the required power budget and the interference strength for different RIS configuration methods. Similar to Fig. \ref{fig:RBD_active_prob}, the required power budgets to achieve $P_d=0.9$ for all the methods increase as the interference becomes stronger. It can be observed that the MF-based method can outperform the ZF-based method when the interference strength is relatively weak. On the contrary, the ZF-based method can surpass the MF-based method and approach the MMSE-based lower bound when the interference strength is relatively strong, which is consistent with the conclusions drawn from the conventional multi-antenna systems. Regarding the passive RIS, we can also see that the MF-based solution no longer approaches the performance of the WMMSE-based solution when the interference is strong. 

Moreover, in the region such that $p_{k}/p_{0}> 16$, the passive RIS configured with the WMMSE-based method can even outperform the active RIS configured with an arbitrary method in terms of the required power budget. The reasons behind this phenomenon are as follows. For the same power budget, although the active RIS is able to amplify the incident signals in addition to adapting the phase shift of REs, it can only support a much smaller number of REs compared to passive RIS because a significant portion of the energy is used to provide amplification capabilities for the active REs. As both greater magnitude of reflection coefficients and larger number of REs are beneficial for improving the sensing performance, the results in Fig. \ref{fig:RBD_PUEA_Ps} exactly demonstrate that increasing the number of REs can achieve more significant performance gain as compared to allocating the power budget to enlarge the amplification factor of REs. For the active RIS, another interesting phenomenon is that the required power budget of each RCM configuration method for $a_{\max}^2=20$dB and $a_{\max}^2=40$dB almost converges to the same value when the interference is strong enough. This can be explained as follows. From Section \ref{subsec:required_power_woPUEA}, we can see that there exists an optimal amplification factor depending on the power of the incident signals and the value of $P_{\rm C}+P_{\rm DC}$. As the power of the incident signals becomes larger due to the stronger interference, the optimal amplification factor will decrease such that the PEAFC is no longer a limiting factor for further reducing the required power budget. 

\begin{figure}[tbp]
    \begin{center}
        \epsfxsize=0.9\linewidth
        \epsffile{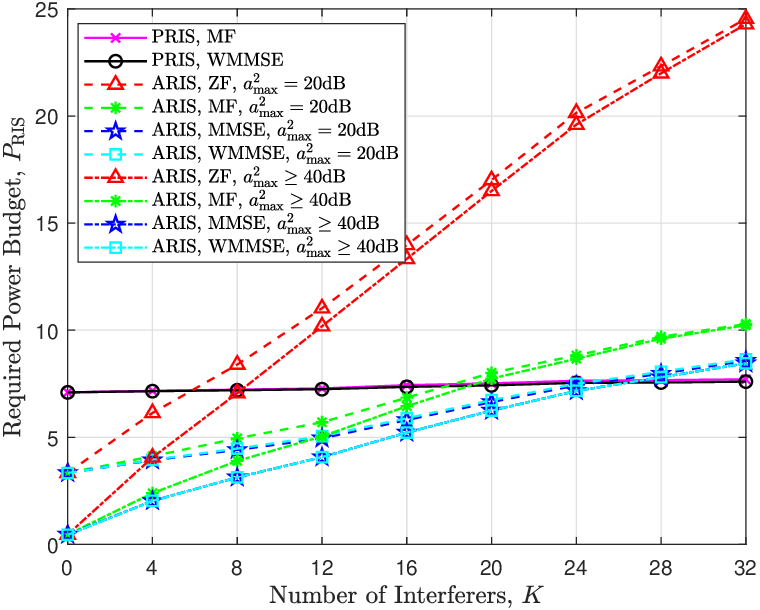}
        \caption{The required power budget to realize a target detection probability of $P_{d}=0.9$ for different numbers of the interferers.}\label{fig:RBD_PUEA_K}
    \end{center}
\end{figure}

In addition, the impacts of the number of interferers are also studied while assuming $\zeta_{1}=\cdots=\zeta_{K}=1$. As shown in Fig. \ref{fig:RBD_PUEA_K}, when there is no interferer ($K=0$), the required power budgets of the active RIS with different configuration methods are the same because these configuration methods produce the same solution in this case. Moreover, for the moderate number of interferers considered in Fig. \ref{fig:RBD_PUEA_K}, the active RIS with MF and MMSE-based configurations can still outperform the passive RIS. On the other hand, the required power budget of the ZF-based method increases more rapidly due to the existence of the budget floor, i.e., $P_{\rm ARIS, ZF}> (K+1)(P_{\rm C}+P_{\rm DC})$. Besides, as $K$ becomes larger, the required power budget of the passive RIS grows less significantly compared to the active RIS, which again demonstrates the capability of massive passive RIS in suppressing the interference. Similar to the case in Fig. \ref{fig:RBD_PUEA_Ps}, when the interference is strong enough due to a large number of interferers, the passive RIS-assisted spectrum sensing system can achieve the target detection probability with a lower power budget. Also, for each active RCM configuration method, the required power budget of $a_{\max}^2=20$dB and that of $a_{\max}^2=40$dB almost converge to the same value in the large $K$ region because the input power also increases along with $K$.

\section{Conclusions}\label{sec:conclusion}

In this paper, we have investigated an active RIS-assisted spectrum sensing system, where the active RIS always forwards the background noise to SU, as it should work all the time to assist the sensing process. The maximum eigenvalue detection method with pre-whitening operations is considered to tackle the correlated background noise incurred by the active RIS and other interferers transmitting in the same band. In particular, we have considered an RCM optimization problem to improve the detection probability given a maximum tolerable false alarm probability and limited sensing time, which is equivalent to maximizing the largest eigenvalue of the population covariance matrix under $\mathcal{H}_1$ with the knowledge of the spiked model from random matrix theory. Then, we have shown that the formulated problem can be equivalently transformed to a weighted mean square error minimization problem using the principle of the well-known WMMSE algorithm, and an iterative optimization approach is proposed to obtain the optimal RCM. In addition, to fairly compare passive RIS and active RIS, we have studied the required power budget of the RIS to achieve a target detection probability under a special case with neglected direct links and LoS RIS-related channels. In addition, we have provided extensive numerical simulations to demonstrate the effectiveness of the WMMSE-based RCM optimization approach. Furthermore, the simulation results have revealed that the active RIS can outperform the passive RIS when the underlying interference within the background noise is relatively weak, whereas the passive RIS performs better in strong interference scenarios because the same power budget can support a vast number of passive REs for interference mitigation.

\bibliographystyle{IEEEtran}
\bibliography{IEEEabrv, ref_ARIS4SS}

\begin{thebibliography}{10}
\providecommand{\url}[1]{#1}
\csname url@samestyle\endcsname
\providecommand{\newblock}{\relax}
\providecommand{\bibinfo}[2]{#2}
\providecommand{\BIBentrySTDinterwordspacing}{\spaceskip=0pt\relax}
\providecommand{\BIBentryALTinterwordstretchfactor}{4}
\providecommand{\BIBentryALTinterwordspacing}{\spaceskip=\fontdimen2\font plus
\BIBentryALTinterwordstretchfactor\fontdimen3\font minus \fontdimen4\font\relax}
\providecommand{\BIBforeignlanguage}[2]{{%
\expandafter\ifx\csname l@#1\endcsname\relax
\typeout{** WARNING: IEEEtran.bst: No hyphenation pattern has been}%
\typeout{** loaded for the language `#1'. Using the pattern for}%
\typeout{** the default language instead.}%
\else
\language=\csname l@#1\endcsname
\fi
#2}}
\providecommand{\BIBdecl}{\relax}
\BIBdecl

\bibitem{ge2023active}
J.~Ge, Y.-C. Liang, and S.~Sun, ``Active {RIS} enhanced spectrum sensing for opportunistic cognitive radio networks,'' in \emph{Proc. {IEEE} Global Commun. Conf. ({GLOBECOM})}, Kuala Lumpur, Malaysia, 2023, pp. 3252--3257.

\bibitem{you2021towards}
X.~You \emph{et~al.}, ``Towards {6G} wireless communication networks: Vision, enabling technologies, and new paradigm shifts,'' \emph{Sci. China Inf. Sci.}, vol.~64, no.~1, pp. 1--74, Nov. 2021.

\bibitem{liang2020dynamic}
Y.-C. Liang, \emph{Dynamic spectrum management: From cognitive radio to blockchain and artificial intelligence}.\hskip 1em plus 0.5em minus 0.4em\relax Springer Nature, 2020.

\bibitem{zeng2010review}
Y.~Zeng, Y.-C. Liang, A.~T. Hoang, and R.~Zhang, ``A review on spectrum sensing for cognitive radio: challenges and solutions,'' \emph{EURASIP J. Adv. Signal Process.}, vol. 2010, pp. 1--15, Jan. 2010.

\bibitem{zeng2009eigenvalue}
Y.~Zeng and Y.-C. Liang, ``Eigenvalue-based spectrum sensing algorithms for cognitive radio,'' \emph{{IEEE} Trans. Commun.}, vol.~57, no.~6, pp. 1784--1793, Jun. 2009.

\bibitem{ghasemi2008spectrum}
A.~Ghasemi and E.~S. Sousa, ``Spectrum sensing in cognitive radio networks: requirements, challenges and design trade-offs,'' \emph{{IEEE} Commun. Mag.}, vol.~46, no.~4, pp. 32--39, Apr. 2008.

\bibitem{zeng2009reliability}
Y.~Zeng, Y.-C. Liang, A.~T. Hoang, and E.~C. Peh, ``Reliability of spectrum sensing under noise and interference uncertainty,'' in \emph{Proc. IEEE Int. Conf. Commun. Workshops}, Dresden, Germany, 2009, pp. 1--5.

\bibitem{lin2020glrt}
M.~Lin, W.~Wang, X.~Hong, and W.~Zhang, ``{GLRT} approach for multi-antenna-based spectrum sensing under interference,'' \emph{{IEEE} Commun. Lett.}, vol.~24, no.~7, pp. 1524--1528, Jul. 2020.

\bibitem{chen2011cooperative}
C.~Chen, H.~Cheng, and Y.-D. Yao, ``Cooperative spectrum sensing in cognitive radio networks in the presence of the primary user emulation attack,'' \emph{{IEEE} Trans. Wireless Commun.}, vol.~10, no.~7, pp. 2135--2141, Jul. 2011.

\bibitem{wu2019towards}
Q.~Wu and R.~Zhang, ``Towards smart and reconfigurable environment: Intelligent reflecting surface aided wireless network,'' \emph{{IEEE} Commun. Mag.}, vol.~58, no.~1, pp. 106--112, Jan. 2019.

\bibitem{basar2019wireless}
E.~Basar, M.~Di~Renzo, J.~De~Rosny, M.~Debbah, M.-S. Alouini, and R.~Zhang, ``Wireless communications through reconfigurable intelligent surfaces,'' \emph{IEEE Access}, vol.~7, pp. 116\,753--116\,773, 2019.

\bibitem{liang2019large}
Y.-C. Liang, R.~Long, Q.~Zhang, J.~Chen, H.~V. Cheng, and H.~Guo, ``Large intelligent surface/antennas ({LISA}): Making reflective radios smart,'' \emph{J. Commun. Inf. Netw.}, vol.~4, no.~2, pp. 40--50, Jun. 2019.

\bibitem{liu2021reconfigurable}
Y.~Liu, X.~Liu, X.~Mu, T.~Hou, J.~Xu, M.~Di~Renzo, and N.~Al-Dhahir, ``Reconfigurable intelligent surfaces: Principles and opportunities,'' \emph{{IEEE} Commun. Surveys Tuts.}, vol.~23, no.~3, pp. 1546--1577, 3rd quarter 2021.

\bibitem{zhou2023assistance}
H.~Zhou, Q.~Zhang, Y.-C. Liang, and Y.~Pei, ``Assistance-transmission tradeoff for {RIS}-assisted symbiotic radios,'' \emph{{IEEE} Trans. Wireless Commun.}, 2023.

\bibitem{zhang2024channel}
Q.~Zhang, H.~Zhou, Y.-C. Liang, W.~Zhang, and H.~V. Poor, ``Channel capacity of {RIS}-assisted symbiotic radios with imperfect knowledge of channels,'' \emph{{IEEE} Trans. Cogn. Commun. Netw.}, 2024.

\bibitem{chen2023transmission}
H.~Chen, R.~Long, and Y.-C. Liang, ``Transmission protocol and beamforming design for ris-assisted symbiotic radio over ofdm carriers,'' in \emph{Proc. {IEEE} Global Commun. Conf. ({GLOBECOM})}, Kuala Lumpur, Malaysia, 2023, pp. 3258--3263.

\bibitem{chen2019intelligent}
J.~Chen, Y.-C. Liang, Y.~Pei, and H.~Guo, ``Intelligent reflecting surface: A programmable wireless environment for physical layer security,'' \emph{IEEE Access}, vol.~7, pp. 82\,599--82\,612, 2019.

\bibitem{jiang2022interference}
T.~Jiang and W.~Yu, ``Interference nulling using reconfigurable intelligent surface,'' \emph{{IEEE} J. Sel. Areas Commun.}, vol.~40, no.~5, pp. 1392--1406, May 2022.

\bibitem{liu2022reconfigurable}
Y.~Liu, X.~Mu, X.~Liu, M.~Di~Renzo, Z.~Ding, and R.~Schober, ``Reconfigurable intelligent surface-aided multi-user networks: Interplay between {NOMA} and {RIS},'' \emph{{IEEE} Wireless Commun.}, vol.~29, no.~2, pp. 169--176, Apr. 2022.

\bibitem{pan2020multicell}
C.~Pan, H.~Ren, K.~Wang, W.~Xu, M.~Elkashlan, A.~Nallanathan, and L.~Hanzo, ``Multicell {MIMO} communications relying on intelligent reflecting surfaces,'' \emph{{IEEE} Trans. Wireless Commun.}, vol.~19, no.~8, pp. 5218--5233, Aug. 2020.

\bibitem{ge2022ris}
J.~Ge, Y.-C. Liang, S.~Li, and Z.~Bai, ``{RIS}-enhanced spectrum sensing: How many reflecting elements are required to achieve a detection probability close to 1?'' \emph{{IEEE} Trans. Wireless Commun.}, no.~10, pp. 8600--8615, Oct. 2022.

\bibitem{wu2021irs}
W.~Wu, Z.~Wang, L.~Yuan, F.~Zhou, F.~Lang, B.~Wang, and Q.~Wu, ``{IRS}-enhanced energy detection for spectrum sensing in cognitive radio networks,'' \emph{{IEEE} Wireless Commun. Lett.}, vol.~10, no.~10, pp. 2254--2258, Oct. 2021.

\bibitem{lin2022intelligent}
S.~Lin, B.~Zheng, F.~Chen, and R.~Zhang, ``Intelligent reflecting surface-aided spectrum sensing for cognitive radio,'' \emph{{IEEE} Wireless Commun. Lett.}, vol.~11, no.~5, pp. 928--932, May 2022.

\bibitem{nasser2022intelligent}
A.~Nasser, H.~A.~H. Hassan, A.~Mansour, K.-C. Yao, and L.~Nuaymi, ``Intelligent reflecting surfaces and spectrum sensing for cognitive radio networks,'' \emph{{IEEE} Trans. Cogn. Commun. Netw.}, vol.~8, no.~3, pp. 1497--1511, Sep. 2022.

\bibitem{wu2023joint}
W.~Wu, Z.~Wang, Y.~Wu, F.~Zhou, B.~Wang, Q.~Wu, and D.~W.~K. Ng, ``Joint sensing and transmission optimization for {IRS}-assisted cognitive radio networks,'' \emph{{IEEE} Trans. Wireless Commun.}, Early Access, 2023.

\bibitem{ge2023ris}
J.~Ge, S.~Wang, C.~Sun, and Y.-C. Liang, ``{RIS}-enhanced cooperative spectrum sensing for opportunistic cognitive radio networks,'' in \emph{2023 IEEE Globecom Workshops (GC Wkshps)}, Kuala Lumpur, Malaysia, 2023, pp. 1427--1432.

\bibitem{long2021active}
R.~Long, Y.-C. Liang, Y.~Pei, and E.~G. Larsson, ``Active reconfigurable intelligent surface-aided wireless communications,'' \emph{{IEEE} Trans. Wireless Commun.}, vol.~20, no.~8, pp. 4962--4975, Aug. 2021.

\bibitem{chen2023active}
Y.~Chen, J.~Wang, and Y.-C. Liang, ``Active reconfigurable intelligent surface-assisted bistatic backscatter communications,'' in \emph{2023 IEEE Globecom Workshops (GC Wkshps)}, Kuala Lumpur, Malaysia, 2023, pp. 1716--1721.

\bibitem{li2023active}
X.~Li, Q.~Zhu, T.~Yu, and Y.~Chen, ``Active {RIS} assisted spectrum sharing: Able to achieve energy-efficient notable detection performance gains,'' \emph{{IEEE} Trans. Veh. Technol.}, vol.~72, no.~9, pp. 11\,668--11\,684, Sep. 2023.

\bibitem{xie2023enhancing}
H.~Xie, B.~Gu, and D.~Li, ``Enhancing spectrum sensing via reconfigurable intelligent surfaces: Passive or active sensing and how many reflecting elements are needed?'' \emph{arXiv preprint arXiv:2306.13874}, 2023.

\bibitem{zeng2008maximum}
Y.~Zeng, C.~L. Koh, and Y.-C. Liang, ``Maximum eigenvalue detection: Theory and application,'' in \emph{Proc. {IEEE} Int. Conf. Commun. ({ICC})}, Beijing, China, 2008, pp. 4160--4164.

\bibitem{ge2021large}
J.~Ge, Y.-C. Liang, Z.~Bai, and G.~Pan, ``Large-dimensional random matrix theory and its applications in deep learning and wireless communications,'' \emph{Random Matrices, Theory Appl.}, vol.~10, no.~04, p. 2230001, 2021.

\bibitem{couillet2011random}
R.~Couillet and M.~Debbah, \emph{Random matrix methods for wireless communications}.\hskip 1em plus 0.5em minus 0.4em\relax Cambridge University Press, 2011.

\bibitem{zhao2023rethinking}
X.~Zhao, S.~Lu, Q.~Shi, and Z.-Q. Luo, ``Rethinking {WMMSE}: Can its complexity scale linearly with the number of {BS} antennas?'' \emph{{IEEE} Trans. Signal Process.}, vol.~71, pp. 433--446, 2023.

\bibitem{shi2015secure}
Q.~Shi, W.~Xu, J.~Wu, E.~Song, and Y.~Wang, ``Secure beamforming for {MIMO} broadcasting with wireless information and power transfer,'' \emph{{IEEE} Trans. Wireless Commun.}, vol.~14, no.~5, pp. 2841--2853, May 2015.

\bibitem{grant2014cvx}
M.~Grant and S.~Boyd, ``{CVX}: {MATLAB} software for disciplined convex programming, version 2.2,'' http://cvxr.com/cvx, Jan. 2020.

\bibitem{jia2021intelligent}
X.~Jia, X.~Zhou, D.~Niyato, and J.~Zhao, ``Intelligent reflecting surface-assisted bistatic backscatter networks: Joint beamforming and reflection design,'' \emph{IEEE Trans. Green Commun. Netw.}, vol.~6, no.~2, pp. 799--814, Jun. 2021.

\bibitem{stevenson2009ieee}
C.~R. Stevenson, G.~Chouinard, Z.~Lei, W.~Hu, S.~J. Shellhammer, and W.~Caldwell, ``{IEEE 802.22}: The first cognitive radio wireless regional area network standard,'' \emph{{IEEE} Commun. Mag.}, vol.~47, no.~1, pp. 130--138, Jan. 2009.

\bibitem{golub2013matrix}
G.~H. Golub and C.~F. Van~Loan, \emph{Matrix computations}.\hskip 1em plus 0.5em minus 0.4em\relax JHU press, 2013.

\bibitem{rappaport1996wireless}
T.~S. Rappaport, \emph{{Wireless Communications: Principles and Practice}}.\hskip 1em plus 0.5em minus 0.4em\relax vol. 2. Upper Saddle River, NJ, USA: Prentice-Hall, 1996.

\end{thebibliography}

\end{document}